\documentclass[journal]{IEEEtran}
\usepackage{graphicx}
\usepackage{amsmath}
\usepackage{mathtools}
\usepackage{amssymb}
\usepackage{hyperref}
\usepackage[noadjust]{cite}
\usepackage{algorithm}
\usepackage[noend]{algpseudocode}
\usepackage{multirow} 
\usepackage{orcidlink}
\DeclareMathOperator*{\argmin}{arg\,min}

\begin{document}
\title{SABER: A Systems Approach to Blur Estimation and Reduction in X-ray Imaging}

\author{K.~Aditya~Mohan\textsuperscript{\orcidlink{0000-0002-0921-6559}},~\IEEEmembership{Senior Member,~IEEE,}
        Robert~M.~Panas\textsuperscript{\orcidlink{0000-0002-7562-0146}},
        and~Jefferson~A.~Cuadra
\thanks{K.~A.~Mohan is with the Computational Engineering Division (CED)
at Lawrence Livermore National Laboratory, Livermore, CA, 94551 USA.
E-mails: mohan3@llnl.gov, adityakadri@gmail.com.}
\thanks{R.~M.~Panas and J.~A.~Cuadra are with the Materials Engineering Division (MED)
at Lawrence Livermore National Laboratory, Livermore, CA, 94551 USA.}
}

%

\maketitle

\begin{abstract}
Blur in X-ray radiographs not only reduces the sharpness of image edges but also reduces the overall contrast.
The effective blur in a radiograph is the combined effect of blur from multiple sources
such as the detector panel, X-ray source spot, and system motion.
In this paper, we use a systems approach to model the point spread function (PSF) of the 
effective radiographic blur as the convolution of multiple PSFs, where each PSF
models one of the various sources of blur.
In particular, we model the combined contribution of X-ray source and detector blurs while assuming negligible contribution from other forms of blur.
Then, we present a numerical optimization algorithm for estimating the source and detector PSFs from multiple radiographs acquired at different X-ray source to object (SOD) and object to detector distances (ODD).
Finally, we computationally reduce blur in radiographs using deblurring
algorithms that use the estimated PSFs from the previous step.
Our approach to estimate and reduce blur is called SABER, which is an acronym for
systems approach to blur estimation and reduction.
\end{abstract}


\section{Introduction}
\begin{figure}[htb]
\begin{center} 
\includegraphics[width=2.3in,keepaspectratio=true]{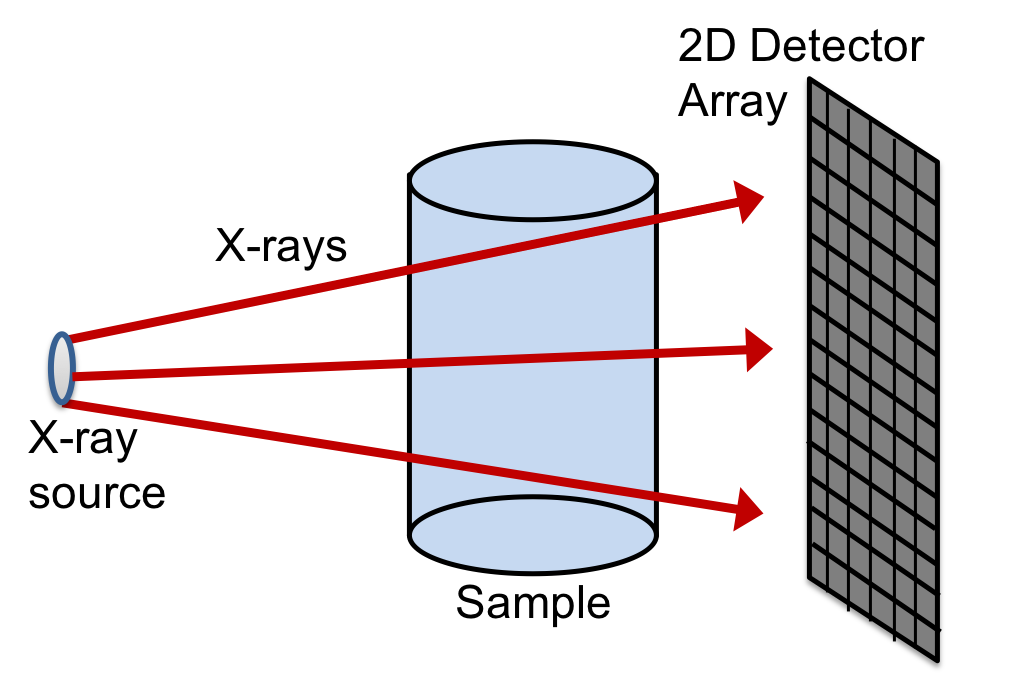}
\end{center}
\caption{\label{fig:expr_setup}
X-ray imaging experimental setup.
An object is exposed to a diverging beam of X-ray radiation 
and the attenuated X-rays emerging from the object are measured by a flat panel of sensor pixels.}
\end{figure}

X-ray imaging systems are widely used for 2D and 3D   
non-destructive characterization and visualization of a wide range of objects.
The ability of X-rays to penetrate deep inside a material makes it a useful tool
 to visualize the interior morphology of objects.
X-ray imaging is very popular in applications such as 
industrial imaging \cite{martz2016book,MohanTIMBIR,GibbsDendrites,Azevedo2016}, 
medical diagnosis \cite{ChestRadGinneken,huda2015radiographic}, and border security \cite{Azevedo2016,Ying2006,Singh2003,Anirudh_2018_CVPR}.
A schematic representation of a typical cone-beam X-ray imaging system is shown in Fig. \ref{fig:expr_setup}.
Here, an object is exposed to a diverging beam of X-ray radiation that is generated by a X-ray source. 
The X-rays get attenuated as they propagate through the object and an image of the attenuated X-ray intensity is recorded
by a detector consisting of a flat-panel 2D array of sensors.

The resolution of an X-ray image, also called a radiograph, is limited by several factors. 
The detector used to record the radiograph imposes a fundamental limit on the resolution by fixing the smallest pixel size.
Also, detector cross-talk where
energy from one sensor pixel leaks into its neighboring sensor pixels \cite{Shefer2013} causes blur and contrast reduction.
Furthermore, the finite non-zero area of the X-ray source aperture manifests as additional blur in the radiograph \cite{LiSourceDeblurTNC,MohanASPEAbs}.
Other causes of blur include motion or vibrations in the sample stage or the imaging equipment. 
Blur is a major detriment in dimensional metrology applications where it is critical to accurately estimate 
the relative physical distances between image features.
In medical applications, blur hinders the 
ability to resolve small features such as tumors that may only be a few pixels in size \cite{de2004high}.  
Hence, mitigating blur is vital to many applications especially when quantitative accuracy is important.
 
In order to quantify radiographic blur,
we use a parametric mathematical model to describe the blurring process.
Several papers address the problem of mathematically modeling numerous types of blur 
such as X-ray source spot blur, detector cross-talk, motion blur, and scatter
\cite{steven2016src,wang2015imaging,LiSourceDeblurTNC}.
A popular strategy is to model
blur as a convolution operation with a certain point spread function (PSF).
Here, the PSF used to model blur is derived using either simulation \cite{WittMCSimBlur} 
or data-driven approaches \cite{FSMSlitPinStar,LinStripBlur,TungsRollbarBlur,CodedApert,SharmaStarPBlurM,JoshiBlurSharpEdge,KeeBlurRemRoundPat}.
Simulation of PSF relies on precise engineering knowledge
of the relevant imaging equipment that may not always be available.
Alternatively, data-driven approaches estimate PSF directly from radiographs
of known well characterized objects. 

Data-driven approaches to blur estimation relies on
calculation of the PSF from radiographs of an object with known composition and shape 
such as a rollbar, slit, pinhole, or other test objects \cite{FSMSlitPinStar,LinStripBlur,TungsRollbarBlur,CodedApert,SharmaStarPBlurM}.
However, these methods only estimate one PSF that
either models the X-ray source blur, the detector blur, or the total effective blur.
They do not address the problem of disentangling and estimating
the PSFs of all the different types of blur that simultaneously affect a radiograph. 
The magnitude of each blur 
 varies depending on the relative placement of the X-ray source, object, and detector. 
The ability to estimate each individual blur PSF will allow us to 
predict the radiograph blur for any spatial configuration of X-ray source, object, and detector
by appropriately recombining the estimated blur PSFs. 

Blur in X-ray radiographs can be reduced either by hardware upgrades
or using computational algorithms to reduce blur after the experiment.
The former approach may not always be feasible due to cost or physical constraints.
Alternatively, deblurring algorithms are a cheaper solution 
to computationally reduce blur in radiographs. 
The estimated blur PSFs 
can be used in a wide variety of deblurring algorithms \cite{WienerHuntDeconv,Orieux2010,TVDeblur,MBIPbook,DeblurringReview,WeinerDeblurNaeem}
 to reduce blur in radiographs.

In this paper, we present an extensible framework called 
systems approach to blur estimation and reduction (SABER)
for modeling, characterizing, and reducing the
various types of blur in a X-ray imaging system.  
We apply a systems approach to model blur
 by expressing the PSF of the effective blur in a radiograph 
as the convolution of multiple blur PSFs with varying origins.
Then, we present a numerical optimization algorithm to disentangle and estimate the individual
 blur PSFs from radiographs of a Tungsten plate rollbar.
 In particular, we focus on the simultaneous estimation of the 
 X-ray source and detector PSFs while assuming negligible motion blur and scatter.  
Preliminary results using this approach was previously published as an extended abstract  \cite{MohanASPEAbs}.
Software that implements SABER is freely available in the form of a
open-source python package called \textit{pysaber} that is documented at the link 
\url{https://pysaber.readthedocs.io/}.



In section \ref{sec:back}, we present the underlying principles and mathematics of X-ray imaging. 
We formulate our blur model in section \ref{sec:blurmodel}
and estimate its parameters using an optimization algorithm in section \ref{sec:blurest}.
In section \ref{sec:debluralgos}, we present two approaches to deblurring radiographs.
Finally, results using real experimental data from a Zeiss Xradia Versa X-ray imaging scanner
are presented in section \ref{sec:results}.

\section{Background}
\label{sec:back}

\begin{figure}[htb]
\begin{center} 
\includegraphics[width=2.5in,keepaspectratio=true]{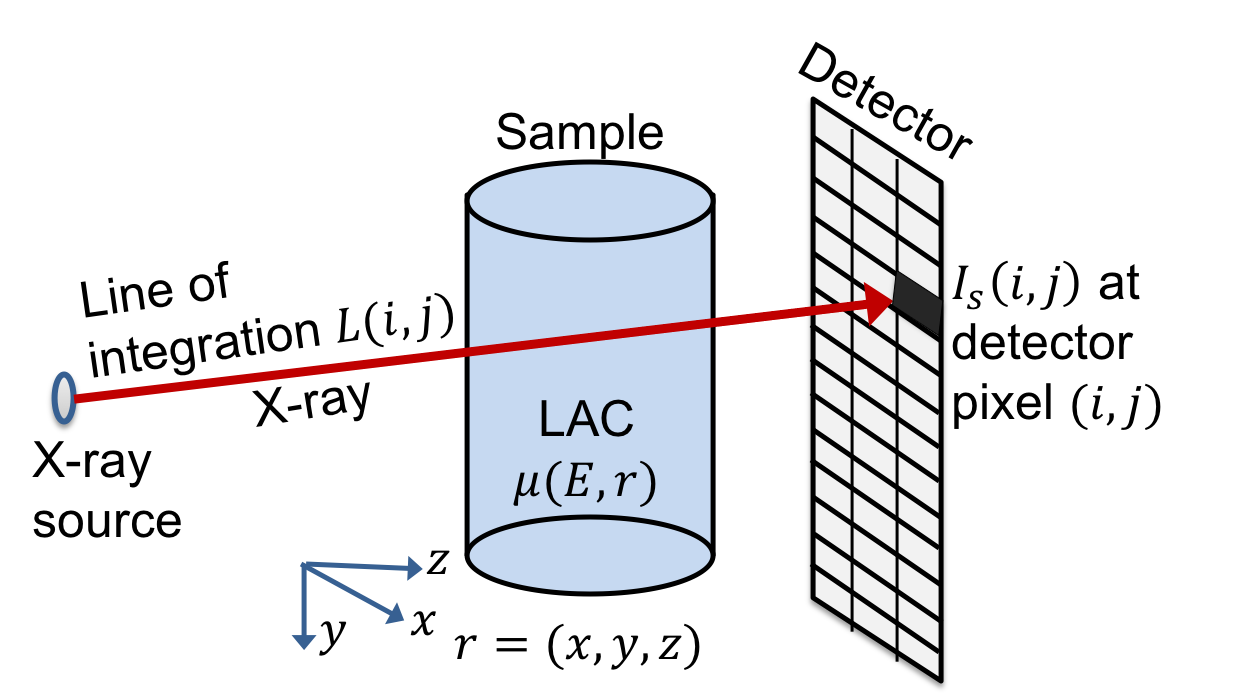}
\end{center}
\caption{\label{fig:line_int}
X-ray propagation model. The detector pixel measurement at location $(i,j)$ is proportional to
the spectrum weighted average of the negative exponential of the integral of the object's LAC $\mu(E,r)$ along the line $L(i,j)$.}
\end{figure}

Beer's law \cite{KakSlaney} is used to express the magnitude of X-ray attenuation 
within an object in terms of its thickness and material properties.
X-ray attenuation is dependent on a material property
called linear attenuation coefficient (LAC) which is a function of 
the object's chemical composition, density, and X-ray energy. 
At detector pixel $(i,j)$ shown in Fig. \ref{fig:line_int}, according to Beer's law, the ratio of the X-ray measurement 
with the object, $I_s(i,j)$, and the X-ray measurement without the object, $I_b(i,j)$,
is given by,
\begin{equation}
\label{eq:polymono}
\frac{I_s(i,j)}{I_b(i,j)} = \int_E S(E) \exp\left(-\int_{L(i,j)}\mu\left(E,r\right)dr\right) dE,
\end{equation}
where $\mu(E,r)$ is the LAC of the object at X-ray energy $E$ and spatial location $r$, 
$L(i,j)$ is the line along which $\mu(E,r)$ is integrated,
and $S(E)$ is the X-ray spectral density such that $\int_E S(E)dE=1$ \cite{Azevedo2016}.
We will call the expression on the right side of the equality in equation \eqref{eq:polymono} 
as the ideal transmission function $\tilde{T}(i,j)$, i.e.,
\begin{equation}
\label{eq:idealtransfunc}
\tilde{T}(i,j) = \int_E S(E) \exp\left(-\int_{L(i,j)}\mu\left(E,r\right)dr\right) dE.
\end{equation}

The relation in equation \eqref{eq:polymono} is
valid in the absence of imaging non-idealities such as noise, blur, and scatter. 
Dark current is one such non-ideal characteristic where the detector 
measurements do not drop to zero when there is no X-ray radiation.
To compensate for this effect,
 detector measurements are made without the X-ray beam and subtracted from measurements with the X-ray beam.
Detector measurements are also corrupted by electronic noise and photon counting noise
that are often modeled as additive Gaussian noise. 
Let $I(i,j)$ denote the normalized measurement defined as the ratio 
 of the dark noise corrected measurements with and without the object.
If $I_d(i,j)$ denotes the measurement at detector pixel $(i,j)$ without X-rays, then,
\begin{equation}
\label{eq:forwmoddarknoise}
I(i,j) = \frac{I_s(i,j)-I_d(i,j)}{I_b(i,j)-I_d(i,j)} = \tilde{T}(i,j) + w(i,j),
\end{equation}
where $w(i,j)$ is Gaussian noise.
In the next section, we will modify equation \eqref{eq:forwmoddarknoise} 
to account for the effects of more non-idealities such as X-ray source blur and detector blur.

The term ``radiograph" will henceforth be used to refer to the normalized radiograph i.e., $I(i,j)$ over all pixel locations $(i,j)$.
Also, ``bright field" is used to refer to $I_b(i,j)$ and ``dark field" is used to refer to $I_d(i,j)$.
 
\section{Formulation of Blur Model}
\label{sec:blurmodel}
The effect of various forms of blur on X-ray radiographs is modeled as a linear space-invariant phenomenon.
The blur model expresses the radiograph $I(i,j)$ as the convolution of a 
transmission function $T(i,j)$ with multiple two dimensional PSFs each of which models a different form of blur. 
In the absence of non-idealities such as scatter or temporal drift in values of $I_b(i,j)$ and $I_d(i,j)$,
the transmission function $T(i,j)$ is equal to the ideal transmission function $\tilde{T}(i,j)$ given in equation \eqref{eq:idealtransfunc}. 


 \begin{figure}[htb]
 \begin{center}
\includegraphics[width=3.5in,keepaspectratio=true]{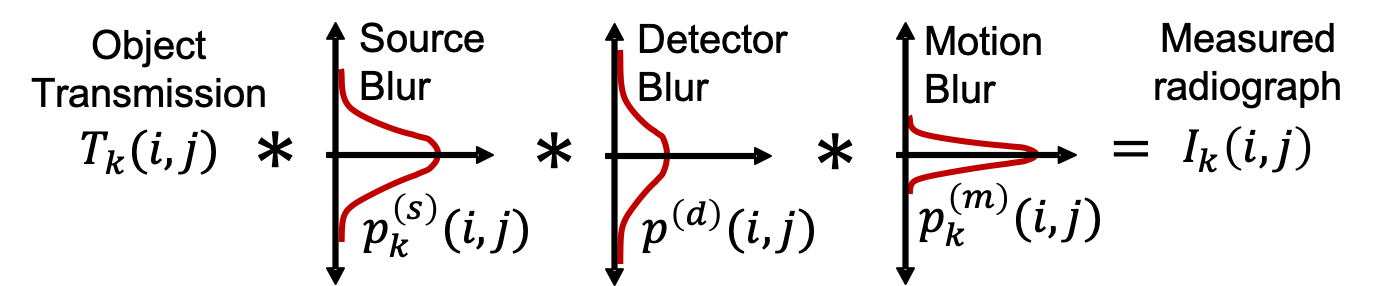}\vspace{-0.1in}
\end{center}
\caption{\label{fig:blur_conv}
The measured radiograph is given by the convolution (denoted by $\ast$) of 
the transmission function with multiple blur PSFs.}
\end{figure}

First, we will consider the effect of blur due to X-ray source, detector, and system motion \cite{hasegawa1990physics}.
The size of the source and motion PSFs are a function of the X-ray source to object distance (SOD) and 
object to detector distance (ODD).
Since detector blur is due to cross-talk between detector pixels, it does not vary with SOD and ODD.
Let $p^{(s)}_k(i,j)$ and $p^{(m)}_k(i,j)$ denote the PSFs of
the X-ray source blur and motion blur respectively on the detector plane at a SOD of $SOD_k$ and ODD of $ODD_k$.
If $I_k(i,j)$ is the radiograph 
and $T_k(i,j)$ is the transmission function at a SOD of $SOD_k$ and ODD of $ODD_k$, then,
\begin{equation}
\label{eq:srcdetblurmodel}
I_k(i,j) = T_k(i,j) \ast p^{(s)}_k(i,j)  \ast p^{(d)}(i,j)  \ast p^{(m)}_k(i,j),
\end{equation}
where $\ast$ denotes 2D convolution (Fig. \ref{fig:blur_conv}).
Based on the analysis in \cite{CuadraASPE2016,CuadraEuspen2017,PanasPCTJournal}, 
we will ignore the effect of motion PSF since
its full width half maximum (FWHM) was determined to be much smaller 
than our detector pixel width. 
Thus, our goal is to estimate the PSFs of X-ray source and detector blur given equation \eqref{eq:srcdetblurmodel}
and radiographs $I_k(i,j)$ at various SOD and ODD.

\subsection{Transmission Function}
Simultaneous estimation of the transmission function, $T_k(i,j)$, and the blur PSFs, $p^{(s)}_k(i,j)$ and $p^{(d)}(i,j)$, is an ill-posed problem.
Hence, we scan a object with known dimensions and chemical composition.
Since our object is known, we can compute the ideal transmission function $\tilde{T}_k(i,j)$ for each scan (indexed by $k$) using equation \eqref{eq:idealtransfunc}.
The transmission function $T_k(i,j)$ is then modeled as an affine transformation
of the ideal transmission function $\tilde{T}_k(i,j)$ such that
\begin{equation}
\label{eq:transvsideal}
T_k(i,j) = l_k + (h_k-l_k)\tilde{T}_k(i,j),
\end{equation}
where $l_k$ and $h_k$ are scalars. 
The parameters $l_k$ and $h_k$ are used to compensate for uniform shifts in the measured X-ray intensity 
 due to non-idealities such as
drift in the values of $I_b(i,j)$ and $I_d(i,j)$ over time, inaccuracies in the calculation 
of $\tilde{T}_k(i,j)$, and  low-frequency effects such as scatter. 
Temporal drifts in $I_b(i,j)$ and $I_d(i,j)$ can occur if 
there is any change in dark current or X-ray source intensity.

\begin{figure}[htb]
\begin{center}
\begin{tabular}{cc}
\includegraphics[width=1.29in,keepaspectratio=true]{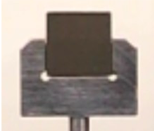}\vspace{-0.1in} &
\includegraphics[width=1.6in,keepaspectratio=true]{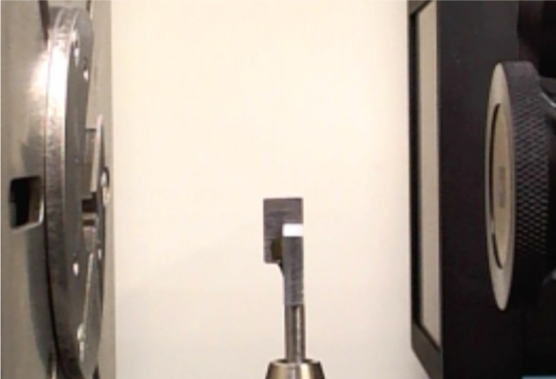} 
\end{tabular}
\end{center}
\caption{\label{fig:rollbarim}
(a) Tungsten plate rollbar mounted in a sample holder. 
(b) Tungsten plate placed in between the X-ray source and detector such that its top rolled edge appears as a horizontal edge in the radiograph.
}
\end{figure}

\begin{figure}[!htb]
\begin{center}
\begin{tabular}{cc}
\hspace{-0.18in} \includegraphics[width=1.7in,keepaspectratio=true]{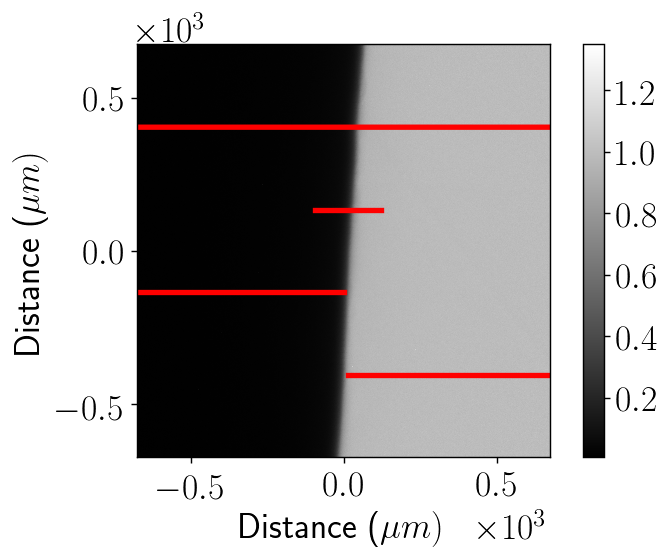} & \hspace{-0.15in}
\includegraphics[width=1.7in,keepaspectratio=true]{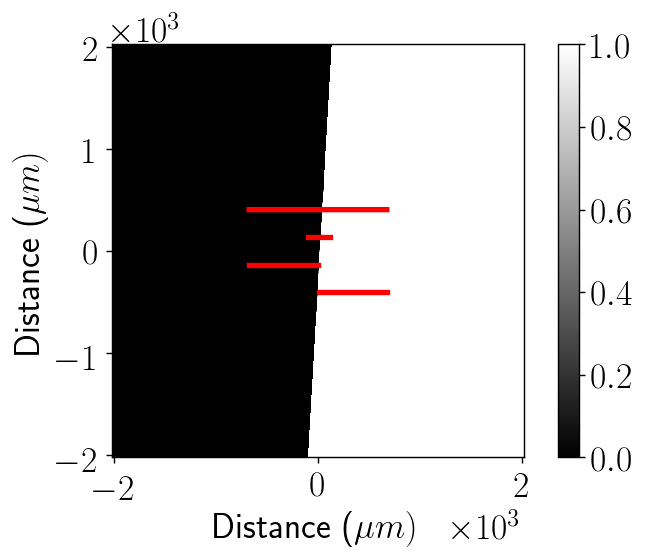} \\
(a) $I_k(i,j)$ & (b) $\tilde{T}_k(i,j)$ \\
\hspace{-0.18in} \includegraphics[width=1.7in,keepaspectratio=true]{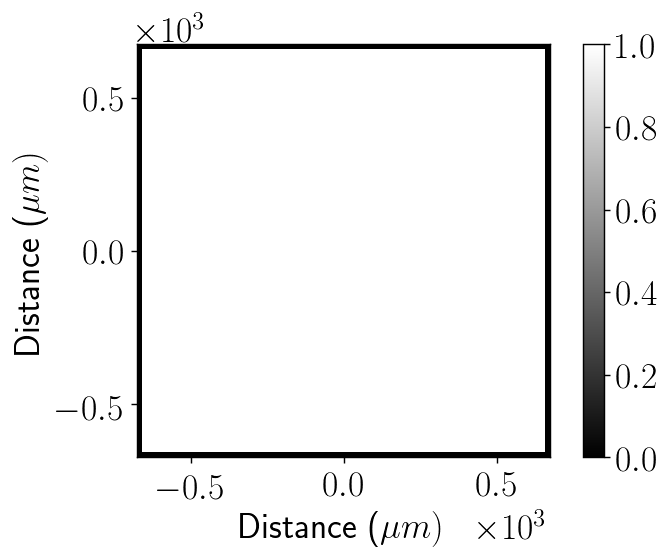} & \hspace{-0.15in}
\includegraphics[width=1.7in,keepaspectratio=true]{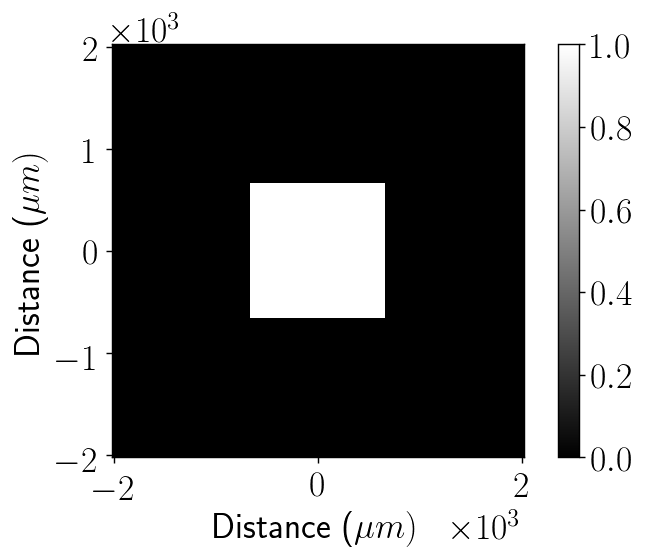} \\
(c) $w(i,j)$  & (d) \\
\hspace{-0.18in} \includegraphics[width=1.5in,keepaspectratio=true]{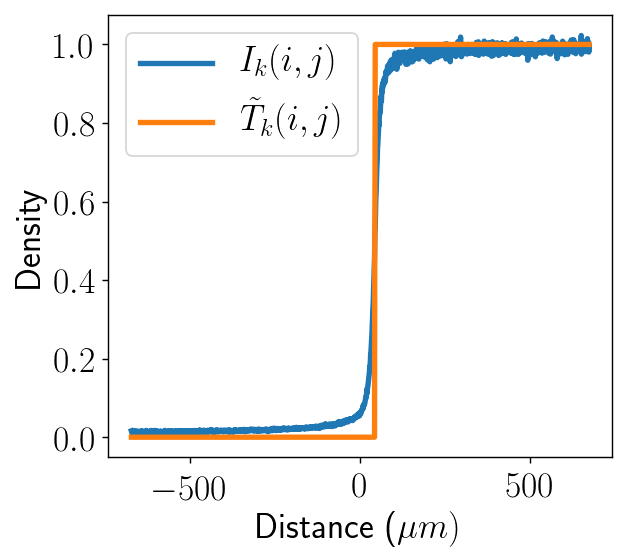} & \hspace{-0.15in}
\includegraphics[width=1.5in,keepaspectratio=true]{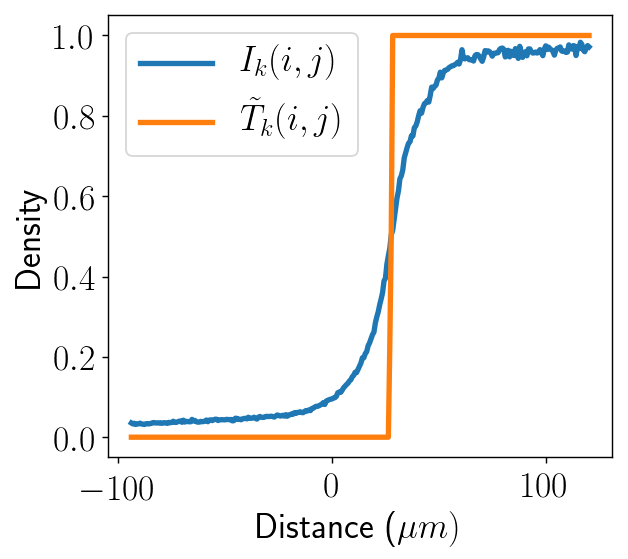} \\
(e) & (f) \\
\hspace{-0.18in} \includegraphics[width=1.5in,keepaspectratio=true]{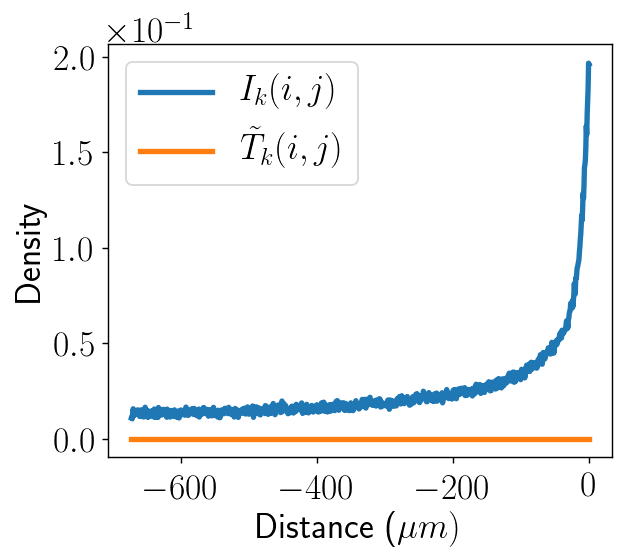} & \hspace{-0.15in}
\includegraphics[width=1.5in,keepaspectratio=true]{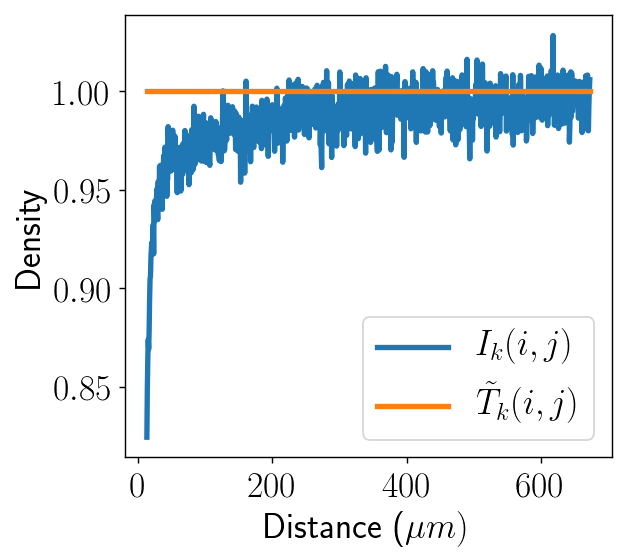} \\
(g) & (h) \\
\end{tabular}
\end{center}
\caption{\label{fig:blurvsidealrad}
(a) Radiograph $I_k(i,j)$ of a Tungsten sharp edge.
(b) Padded ideal transmission function $\tilde{T}_k(i,j)$ that is derived from (a).
The bright regions in (c) and (d) indicate the regions of (a) and (b) respectively 
that is included in the optimization of equation \eqref{eq:blurmopt}. 
The origin $(0,0)$ is at the center of all images (a-d).
The bright regions in (c) and (d) have the same number of pixels and same pixel width.
The width of the image in (d) is greater than (c) due to additional padding in (d).
(e), (f), (g), and (h) are line profile plots along the red colored lines in images (a) and (b). }
\end{figure}

\subsubsection{Calibration Data}
For our experiments, we fabricated a Tungsten plate with a sharp edge
and uniform thickness that is sufficient to block all incoming X-rays (Fig. \ref{fig:rollbarim}). 
The plate is then exposed to X-rays such that the sharp edge
passes close to the center of the radiograph as shown in Fig. \ref{fig:blurvsidealrad} (a).
The plane of the Tungsten plate is aligned such that it is perpendicular to the direction of X-ray propagation.
Since perfect alignment is very challenging, the sharp edges are rolled \cite{TungsRollbarBlur} 
to one meter radius of curvature, which permits alignment errors of up to one degree. 
Fig. \ref{fig:blurvsidealrad} (b) shows the ideal transmission function for the radiograph in Fig. \ref{fig:blurvsidealrad} (a).
The procedure used to compute the ideal transmission function is presented in Appendix \ref{app:itransfunc}.

\subsubsection{Initialization of $l_k$ and $h_k$}
\label{sec:initlkhk}
Due to scatter and variations in $I_b(i,j)$ and $I_d(i,j)$, $T_k(i,j)$ is not the same as $\tilde{T}_k(i,j)$.
The values of $l_k$ and $h_k$ that determine the relation between $T_k(i,j)$ and $\tilde{T}_k(i,j)$ in equation \eqref{eq:transvsideal} are jointly 
estimated during the blur estimation procedure that will be presented in the next section. 
However, to aid this procedure, we have to supply initial estimates for $l_k,h_k$ 
that are determined by finding the best values for $l_k,h_k$ that solve the following over-determined set of linear equations,
\begin{multline}
\label{eq:ransacreg}
I_k(i,j) = l_k + (h_k-l_k)\tilde{T}_k(i,j),
\text{ where } 
w(i,j) = 1.
\end{multline}
Here, $w(i,j)$ is the weight term that is $0$ for pixels that are close to the boundaries of the image $I_k(i,j)$ and $1$ elsewhere (Fig. \eqref{fig:blurvsidealrad} (c)).
The padded region of $\tilde{T}(i,j)$ is also excluded from consideration during parameter estimation (Fig. \eqref{fig:blurvsidealrad} (d)).
Equation \eqref{eq:ransacreg} is solved using \textit{RANSAC} regression \cite{ransacref,scikit-learn}.
We used the \textit{scikit-learn} \cite{scikit-learn} implementation of \textit{RANSAC}\footnote{\textit{RANSAC} Parameters: Minimum number of randomly chosen samples was $10$. For a data sample to be classified as an inlier, the maximum residual threshold was $0.1$.} in this paper.

\subsection{Source Blur}
The PSF of source blur is mathematically modeled using a 2D density function. 
This function is parameterized by two scale parameters $s_{sx}$ and $s_{sy}$
that are a measure of the spatial width of the PSF 
along the $x-$axis and $y-$axis in the plane of the X-ray source.
If $\Delta$ denotes the width of each pixel,
the PSF of source blur in the plane of the X-ray source is modeled as,
\begin{equation}
\label{eq:srcpsf}
p^{(s)}\left(i,j\right)= \frac{1}{Z_s} \exp\left( -\Delta^r\left(i^2 s^2_{sx} + j^2 s^2_{sy} \right)^{\frac{r}{2}} \right),
\end{equation}
where $r\geq 1$ and $Z_s$ is the normalizing constant given by,
\begin{equation}
\label{eq:srcnorm}
Z_s = \sum_i\sum_j\exp\left( -\Delta^r\left(i^2 s^2_{sx} + j^2 s^2_{sy} \right)^{\frac{r}{2}} \right).
\end{equation}
The constant $Z_s$ ensures that $p^{(s)}\left(i,j\right)$ sums to one when summed over all $(i,j)$.
In equation \eqref{eq:srcpsf}, a Gaussian density function is obtained by setting $r=2$ and an 
exponential density function is obtained when $r=1$.
The scale parameters are related to the full width half maximums (FWHM) along 
the $x-$axis as $W_{sx}=2\log(2)^{\frac{1}{r}}/s_{sx}$ and along the $y-$axis as $W_{sy}=2\log(2)^{\frac{1}{r}}/s_{sy}$.
By definition, FWHM $W_{sx}$ is the distance between two points along the $x-$axis
where the PSF drops to half of its maximum value and FWHM $W_{sy}$ is the corresponding distance
along the $y-$axis.

\begin{figure}[htb]
\begin{center} 
\includegraphics[width=2.3in,keepaspectratio=true]{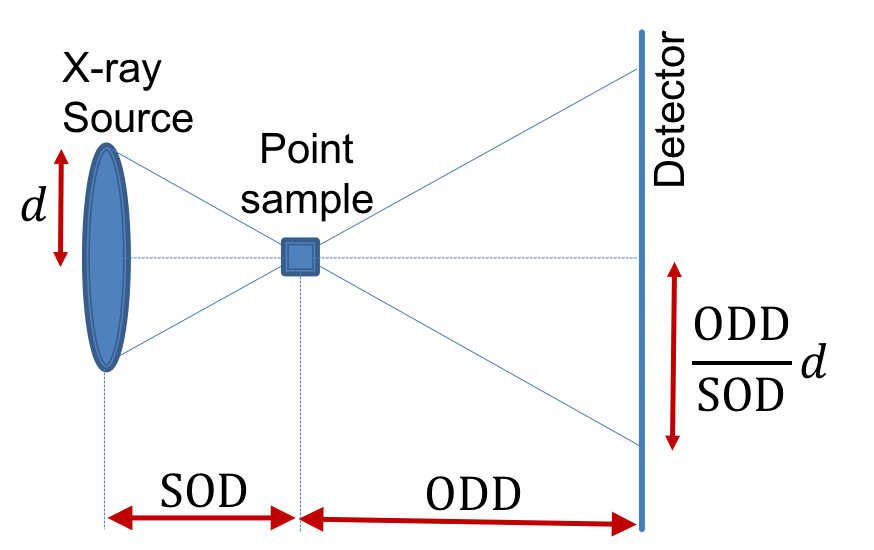}
\end{center}
\caption{\label{fig:source_blur_expr}
The width of source blur on the detector plane is directly proportional to the object to detector distance (ODD) 
and inversely proportional to the source to object distance (SOD).}
\end{figure}

At the detector plane, the FWHM of source blur is scaled by a factor of ODD/SOD.
The change in FWHM of source blur with varying ODD and SOD is depicted in Fig. \ref{fig:source_blur_expr}.
For the $k^{th}$ radiograph, the PSF of source blur on the detector plane is given by,
\begin{equation}
\label{eq:srcdetpsf}
p^{(s)}_k(i,j)=\frac{1}{Z_{s,k}}\exp\left( -\Delta^r \frac{SOD^r_k}{ODD^r_k}\left(i^2 s^2_{sx} +j^2 s^2_{sy} \right)^{\frac{r}{2}} \right),
\end{equation}
where $Z_{s,k}$ is the normalizing constant given by,
\begin{equation}
\label{eq:srcdetpsfnorm}
Z_{s,k} = \sum_i\sum_j\exp\left( -\Delta^r \frac{SOD^r_k}{ODD^r_k}\left(i^2 s^2_{sx} +j^2 s^2_{sy} \right)^{\frac{r}{2}} \right).
\end{equation}

\subsection{Detector Blur}
The detector blur is modeled using a 
mixture of two density functions with scale parameters $s_{d1}$ and $s_{d2}$. 
If $p^{(d)}(i,j)$ denotes the PSF of detector blur, then,
\begin{multline}
\label{eq:detpsf}
p^{(d)} (i,j) = q \frac{1}{Z_{d1}} \exp\left(-s^r_{d1} \Delta^r \left(i^2+j^2\right)^{\frac{r}{2}} \right)\\
+(1-q)  \frac{1}{Z_{d2}} \exp\left(-s^r_{d2} \Delta^r \left(i^2+j^2\right)^{\frac{r}{2}} \right)
\end{multline}
where $q$ is the mixture parameter and $Z_{d1},Z_{d2}$ are normalizing constants such that
\begin{align}
\label{eq:detpsfnorm1}
Z_{d1} & = \sum_i \sum_j \exp\left(-s^r_{d1} \Delta^r \left(i^2+j^2\right)^{\frac{r}{2}} \right), \\
\label{eq:detpsfnorm2}
Z_{d2} & = \sum_i \sum_j \exp\left(-s^r_{d2} \Delta^r \left(i^2+j^2\right)^{\frac{r}{2}} \right).
\end{align}
Also, $q$ and $(1-q)$ are parameters that function as weights for the two density functions. 
The scale parameters are related to the FWHMs of the two
exponential functions in equation \eqref{eq:detpsf} as
 $W_{d1}=2\log(2)^{\frac{1}{r}}/s_{d1}$ and  $W_{d2}=2\log(2)^{\frac{1}{r}}/s_{d2}$.
 A Gaussian mixture model for detector blur is obtained by setting $r=2$ and an exponential mixture model is obtained when $r=1$.
 
In our experiments, we expect the detector PSF to be dominated
by the first exponential with a large weight of $q\approx 0.9$ and a very small FWHM $W_{d1}$ that spans only a few pixels.
The second exponential is expected to have a smaller weight of $1-q\approx 0.1$ 
and a very large FWHM $W_{d2}$ that spans several hundreds of pixels.
Depending on the X-ray imaging system, the values for $W_{d1},W_{d2}$ and $q$ may change but equation
\eqref{eq:detpsf} is still expected to be a good model for detector PSF.

\subsection{Motion Blur}
An additional source of blur is the relative motion of all components in the X-ray system.  
If these components move during the data acquisition period, 
then they will result in motion of the object's image on the detector surface.  
Such motion will distribute the apparent location of an edge or feature over a range of locations, 
appearing to blur it in a manner that can be captured with a motion PSF that is convolved with the image.  
The system motion is calculated via a geometric analysis of the relative locations of components, propagated to the detector screen. 
This is done using an uncertainty propagation model presented in
 \cite{CuadraASPE2016,CuadraEuspen2017,PanasPCTJournal}. 
Motion blur is denoted by $p^{(m)}_k(i,j)$ and is typically modeled as a Gaussian density function.
This form of blur also depends on the relative positions of X-ray source, object, and detector.
However, we will effectively ignore the motion PSF by modeling it as a delta function since
its FWHM was determined to be much smaller 
than the detector pixel width in our experiments. 
Thus, the form of $p^{(m)}_k(i,j)$ is given by,
\begin{equation}
\label{eq:motblurdelta}
p^{(m)}_k(i,j) = 
\begin{cases}
1, & \text{ if } i=0,j=0\\
0, & \text{ otherwise}
\end{cases}.
\end{equation}

\section{Estimation of Blur Model}
\label{sec:blurest}
Blur model estimation is the process of estimating
all the parameters $s_{sx},s_{sy},s_{d1},s_{d2},$ and $q$ of the PSFs 
given known values for $I_k(i,j)$ and $\tilde{T}_k(i,j)$ in equations \eqref{eq:srcdetblurmodel} and \eqref{eq:transvsideal}.
We will only estimate $p^{(s)}_k(i,j)$
and $p^{(d)}(i,j)$ in equation \eqref{eq:srcdetblurmodel} since $p^{(m)}_k(i,j)$ is assumed to be a 
constant delta function. 
By constraining $p^{(s)}_k(i,j)$ to take the shape of the density function
in equation \eqref{eq:srcpsf} and $p^{(d)}(i,j)$ to take the shape of the mixture density function
in equation \eqref{eq:detpsf}, blur estimation reduces to the problem of estimating the parameters
$s_{sx}$, $s_{sy}$, $s_{d1}$, $s_{d2}$, and $q$.
For every $k^{th}$ radiograph, the parameters $l_k$ and $h_k$ in equation \eqref{eq:transvsideal} are also treated as unknowns
and jointly estimated along with the PSF parameters.

Note that the form of source PSF as evaluated on the detector plane
given by equation \eqref{eq:srcdetpsf} depends on the ratio of the object to detector distance $ODD_k$ 
and the source to object distance $SOD_k$.
Thus, while the amount of source blur in the radiograph $I_k(i,j)$ is a function of $ODD_k$ and $SOD_k$,
 the detector blur does not change with $ODD_k$ and $SOD_k$.
To estimate both source and detector PSF parameters,
it is necessary to acquire radiographs at a minimum of two different values of $ODD_k/SOD_k$.
Also, since the source PSF has two scale parameters modeling the width
along $x-$axis and $y-$axis, 
we need radiographs for cases when the Tungsten edge is horizontal and vertical.

We use numerical optimization to estimate all
blur and transmission parameters given radiographs at different values
of ODD and SOD.
The parameters are estimated by solving,
\begin{equation}
\label{eq:blurmopt}
\left(\begin{matrix}\hat{s}_{sx},\hat{s}_{sy},\hat{s}_{d1},\hat{s}_{d2},\hat{q}, 
\\ \hat{l}_1,\cdots,\hat{l}_K,\hat{h}_1,\cdots,\hat{h}_K \end{matrix}\right)  = 
\hspace{-0.3in}\argmin_{\begin{matrix}s_{sx},s_{sy}, s_{d1},s_{d2},q \\ l_1,\cdots,l_K, h_1,\cdots,h_K\end{matrix}} \sum_{k=1}^K E_k 
\end{equation}
\vspace{-0.2in}
 \begin{multline}
 \label{eq:optconstr}
 \text{where } E_k=0.5\sum_{i,j}w_k(i,j)\left(I_k(i,j)-\right. \\
\left. T_k (i,j) \ast p^{(s)}_{k}(i,j) \ast p^{(d)}(i,j) \ast p^{(m)}_k(i,j)\right)^2,\\
\text{subject to }s_{sx}\geq 0, s_{sy}\geq 0, s_{d1}\geq 0, s_{d2}\geq 0, \\ q_{low} \leq q\leq 1,
-1\leq l_k\leq 0.5, 0.5 \leq h_k \leq 2,
\end{multline}
where $q_{low}=0.8$ and $K$ is the total number of radiographs.
The parameters
 $\hat{s}_{sx}$, $\hat{s}_{sy}$, $\hat{s}_{d1}$, $\hat{s}_{d2}$, $\hat{q}$, $\hat{l}_k$,
and $\hat{h}_k,\forall k$ are the 
estimated values of $s_{sx}$, $s_{sy}$, $s_{d1}$, $s_{d2}$, $q$, $l_k$,
and $h_k,\forall k$ respectively. The lower bound of $0.8$ for $q$ ensures that 
the first exponential in equation \eqref{eq:detpsf} is chosen to have the largest weight.

\begin{algorithm}
\caption{Blur Estimation Algorithm}\label{alg:blurest}
\begin{algorithmic}[1]
\State Estimation of $s_{sx}$ and $s_{sy}$ from $K/2$ number of radiographs 
with the highest $ODD_k/SOD_k$. Let $\Omega_s$ be the set of all indices 
of these radiographs. Parameters $l_k$ and $h_k$ for all $k\in \Omega_s$ are set using the procedure in section \ref{sec:initlkhk}.
\begin{equation}
\label{eq:algopt1}
\left(\hat{s}_{sx},\hat{s}_{sy}\right) 
= \argmin_{s_{sx}\geq 0,s_{sy} \geq 0} \sum_{k\in \Omega_s} E_k.
\end{equation}
\State Estimation of $s_{d1}$, $s_{d2}$, and $q$ from $K/2$ number of radiographs 
with the lowest $ODD_k/SOD_k$. Let $\Omega_d$ be the set of all indices 
 of these radiographs.
Parameters $l_k$ and $h_k$ for all $k\in \Omega_d$ are set using the procedure in section \ref{sec:initlkhk}.
\begin{equation}
\label{eq:algopt2}
\left(\hat{s}_{d1},\hat{s}_{d2},\hat{q}\right) 
= \argmin_{s_{d1}\geq 0,s_{d2}\geq 0,q_{low}\leq q\leq 1} \sum_{k\in \Omega_d} E_k
\end{equation}
\State Estimation of all parameters $s_{sx}$, $s_{sy}$, $s_{d1}$, $s_{d2}$, $q$, $l_k$, and $h_k$. 
$s_{sx}$ and $s_{sy}$ are initialized with the estimated values $\hat{s}_{sx}$ and $\hat{s}_{sy}$ from step 1. 
$s_{d1}$, $s_{d2}$, and $q$ are initialized with the estimated values $\hat{s}_{d1},\hat{s}_{d2},$ and $\hat{q}$ from step 2. 
$l_k$ and $h_k$ are initialized using the procedure in section \ref{sec:initlkhk}.
\begin{multline}
\label{eq:algopt3}
\left(\begin{matrix}\hat{s}_{sx},\hat{s}_{sy},\hat{s}_{d1},\hat{s}_{d2},\hat{q}, 
\\ \hat{l}_1,\cdots,\hat{l}_K,\hat{h}_1,\cdots,\hat{h}_K \end{matrix}\right)  = 
\hspace{-0.3in}\argmin_{\begin{matrix}s_{sx},s_{sy}, s_{d1},s_{d2},q \\ l_1,\cdots,l_K, h_1,\cdots,h_K\end{matrix}} \sum_{k=1}^K E_k \\
\text{subject to the constraints in equation \eqref{eq:optconstr}}. 
\end{multline}
\end{algorithmic}
\end{algorithm}

We solve the minimization problem in equation \eqref{eq:blurmopt} using the L-BFGS-B algorithm \cite{LBFGSBTheory}.
L-BFGS-B is a quasi-Newton method for solving optimization problems 
and requires information of the gradient of the objective function, $\sum_{k=1}^K E_k$, with 
respect to all the variables being optimized.
It also supports bound constraints on each of the variables that are optimized. 
However, since the optimization problem in equation \eqref{eq:blurmopt} is non-convex, 
the optimized variables may be stuck in a local minima or a saddle point. 
Thus, good initialization of each variable is necessary  to ensure reliable convergence to a solution that best fits the measured radiographs.
Our approach to solving equation \eqref{eq:blurmopt} is outlined in algorithm \ref{alg:blurest}.
We use the L-BFGS-B implementation in the python package \textit{scipy} \cite{scipylib,LBFGSBFortran}.
The gradients that must be supplied to the L-BFGS-B algorithm are derived in Appendix \ref{app:objgrad}.

In algorithm \ref{alg:blurest}, steps $1$ and $2$ are used to produce 
good initial estimates of the source and detector PSF parameters 
for initializing the optimization in step $3$.
In step $1$, only source PSF parameters are estimated 
from half of all radiographs with the highest $ODD_k/SOD_k$
where source blur is significant if not dominant.
In step $2$, only detector PSF parameters are estimated 
from half of all radiographs with the lowest $ODD_k/SOD_k$
where detector blur is significant if not dominant.
In step $3$, all parameters from the source PSF, detector PSF,
and transmission function are jointly optimized to arrive at the final estimates. 

\section{Deblurring Algorithms}
\label{sec:debluralgos}
Deblurring is the process of reducing blur in radiographs using computational algorithms.
We focus on using Wiener filter \cite{Orieux2010,DeblurringReview} and 
regularized least squares deconvolution \cite{MBIPbook,TVDeblur,DeblurringReview} for deblurring radiographs.
These algorithms take as input the convolution of all PSFs given by, 
\begin{equation}
\label{eq:allpsfconv}
\hat{p}_k\left(i,j\right)=\hat{p}^{(s)}_k(i,j) \ast \hat{p}^{(d)}(i,j) \ast \hat{p}^{(m)}_k(i,j),
\end{equation}
which is a function of the source to object distance $SOD_k$
and object to detector distance $ODD_k$ of the input radiograph $I_k(i,j)$.
In equation \eqref{eq:allpsfconv}, $\hat{p}^{(s)}_k(i,j)$ and $\hat{p}^{(d)}(i,j)$ are obtained by substituting the  
estimated values $\hat{s}_{sx},\hat{s}_{sy},\hat{s}_{d1}, \hat{s}_{d2},$ and $\hat{q}$ from step 3 
of algorithm \ref{alg:blurest} in equations \eqref{eq:srcdetpsf} and \eqref{eq:detpsf}.
The PSF $\hat{p}^{(m)}_k(i,j)$ is assumed to be a discrete delta function (equation \eqref{eq:motblurdelta}) since 
FWHM of motion blur was determined to be much smaller than the pixel width
using the analysis in  \cite{CuadraASPE2016,CuadraEuspen2017,PanasPCTJournal}. 

\subsection{Wiener Filter}
Wiener filter \cite{Orieux2010,DeblurringReview} reduces blur by deconvolving the convolution of all PSFs, 
$\hat{p}_k\left(i,j\right)$ in equation \eqref{eq:allpsfconv}, from the blurred radiograph $I_k(i,j)$.
Deconvolution is implemented in Fourier space by dividing the Fourier transform
of the radiograph with the Fourier transform of the PSFs. 
To reduce noise, regularization is used to enforce smoothness.
We use the function \textit{skimage.restoration.wiener} in the python package \textit{scikit-image} 
\cite{scikit-image} that implements the method in \cite{Orieux2010}.


\subsection{Regularized Least Squares Deconvolution (RLSD)}
In RLSD \cite{MBIPbook,bouman1993generalized,DeblurringReview,TVDeblur}, we solve the following optimization problem to deblur a radiograph 
$I_k(i,j)$,
\begin{multline}
\label{eq:debluropt}
\hspace{-0.15in}\hat{T}_k(i,j)=\\
\argmin_{T_k(i,j),\forall i,j}\left\lbrace \sum_{i,j}w_k(i,j)\left(I_k (i,j) - 
 T_k(i,j) \ast \hat{p}_k\left(i,j\right) \right)^2 \right. \vspace{-0.2in}\\
\left. +\beta \hspace{-0.25in} \sum_{((i,j),(m,n))\in \mathcal{N}} \hspace{-0.25in} \tilde{w}(i,j,m,n) |T_k(i,j)-T_k(m,n)|^{1.2} \right\rbrace,
\end{multline}
where $\hat{T}_k(i,j)$ is the deblurred radiograph, 
$\beta$ is the regularization parameter, 
and $\mathcal{N}$ is the set of all pairs of neighboring pixel indices 
i.e., $((i,j),(m,n))\in \mathcal{N}$ if pixel $(m,n)$ lies within a $3\times 3$ neighborhood of pixel $(i,j)$.
The regularization weight parameter $\tilde{w}(i,j,m,n) \propto \frac{1}{\sqrt{(i-m)^2+(j-n)^2}}$ is inversely proportional 
to the distance between neighboring pixels
and normalized such that $\sum_{(m,n)\in \mathcal{N}_{(i,j)}} \tilde{w}(i,j,m,n)=1$,
where $\mathcal{N}_{(i,j)}$ is the set of all
voxel indices that are neighbors to voxel index $(i,j)$.
For simplicity, the weight parameter $w_k(i,j)$ is chosen to be $1$ for all $(i,j)$ in our experiments.
Ideally, $w_k(i,j)$ should be set such that it is inversely proportional to the variance of noise in $I_k(i,j)$.

The regularization function in equation \eqref{eq:debluropt}
 enforces smoothness in $T_k(i,j)$ by penalizing the difference in values between 
neighboring pixels \cite{bouman1993generalized,Mohan4DCT}.
The optimization problem in equation \eqref{eq:debluropt} is solved
using the L-BFGS-B algorithm \cite{LBFGSBTheory}.

\section{Results}
\label{sec:results}
In this section, we will estimate the blur model for a Zeiss Xradia 510 Versa X-ray imaging system. 
We will compare the efficacy of our blur model against conventional approaches 
using quantitative metrics. 
After estimating the blur model, we use Wiener filtering and RLSD (described in section \ref{sec:debluralgos})
to deblur radiographs.

The Versa is a commercial micro-CT system that consists of a transmissive X-ray
tube with a Tungsten target anode that exhibits bremsstrahlung X-ray spectral characteristics. 
The accelerating voltage was selected to be $160 kV$ and the tube current was $62.5\mu A$.
This resulted in an average flux of $1.37 \times 10^{11} counts\, mA^{-1} nstr^{-1} sec^{-1}$. 
The detector consists of an optically coupled Thallium-doped Cesium 
Iodide scintillator and a $5 MP$ CCD with a $13.5\mu m$ pixel width, $16$-bit depth, 
and a maximum dark current noise of about $12 \, counts\, sec^{-1}$. 
Bright field images (radiographs without object) $I_b(i,j)$ and dark field images (radiographs with X-rays off) $I_d(i,j)$
were acquired to appropriately normalize each radiograph image of the object using equation \eqref{eq:forwmoddarknoise}.
All radiographs were acquired using a $20\times$ magnification lens resulting in an effective pixel width of $\Delta=0.675\mu m$.

\subsection{Blur PSF Parameter Estimation}
Radiographs of a Tungsten plate rollbar are used to estimate the blur model parameters
 $s_{sx},s_{sy},s_{d1},s_{d2}$, and $q$ by solving the optimization problem in equation \eqref{eq:blurmopt}.
 The source to detector distance (SDD) for all radiographs was fixed at $71mm$.
First, the Tungsten plate edge is oriented in the vertical direction (Fig. \ref{fig:blurvsidealrad} (a)) and radiographs are acquired
at source to object distances (SOD) of $13mm$, $24.8mm$, $37.5mm$, $50.3mm$, and $60mm$.
Next, the edge is oriented in the horizontal direction and 
radiographs are acquired at SOD of $12mm$, $24.8mm$, $37.5mm$, $50.3mm$, and $65.3mm$.
Since SDD is $71mm$, the object to detector distance (ODD) for each radiograph is given by ($71-$SOD)$mm$. 
For each radiograph, the Tungsten plate is oriented such that its edge is slightly tilted away from the horizontal or vertical directions
as shown in Fig. \ref{fig:blurvsidealrad} (a).
This is done to ensure that the Tungsten edge is never  parallel to a row or column of detector pixels.

\begin{table}[b!]
\begin{center}
\caption{FWHMs $W_{sx}, W_{sy}$ of the X-ray source PSF 
 estimated using one set of horizontal and vertical edge radiographs.}
\label{tab:sourceonlypar}
\begin{tabular}{|c|c|c|c|}
\hline
horizontal & vertical & $W_{sx}$ & $W_{sy}$ \\
SOD ($mm$) &  SOD ($mm$) & ($\mu m$) & ($\mu m$) \\ \hline
12.0 & 13.0 & 3.2 & 3.6 \\\hline
24.8 & 24.8 & 3.6 & 3.9 \\\hline
37.5 & 37.5 & 4.5 & 4.6 \\\hline
50.3 & 50.3 & 6.9 & 6.9 \\\hline
65.3 & 60.0 & 13.0 & 25.9 \\\hline
\end{tabular}
\end{center}
\end{table}

\begin{table}[h!]
\begin{center}
\caption{Parameters $q$, $W_{d1}$, and $W_{d2}$ of the detector PSF estimated
using one set of horizontal and vertical edge radiographs.}
\label{tab:detonlypar}
\begin{tabular}{|c|c|c|c|c|}
\hline
horizontal & vertical & $W_{d1}$ & $W_{d2}$ & \multirow{2}{*}{$q$} \\
SOD ($mm$) &  SOD ($mm$) & ($\mu m$) & ($\mu m$) & \\ \hline
12.0 & 13.0 & 12.6 & 130.6 & 0.89 \\\hline
24.8 & 24.8 & 5.7 & 134.9 & 0.91 \\\hline
37.5 & 37.5 & 3.4 & 158.3 & 0.92 \\\hline
50.3 & 50.3 & 2.3 & 127.0 & 0.92 \\\hline
65.3 & 60.0 & 1.9 & 118.3 & 0.93 \\\hline
\end{tabular}
\end{center}
\end{table}

\begin{table}[h!]
\begin{center}
\caption{X-ray source PSF and detector PSF parameters estimated using two sets of 
horizontal and vertical edge radiographs.}
\label{tab:srcdetpar}
\begin{tabular}{|c|c|c|c|c|c|c|}
\hline
horizontal & vertical & $W_{sx}$ & $W_{sy}$ & $W_{d1}$ & $W_{d2}$ & \multirow{2}{*}{$q$} \\
SOD ($mm$) &  SOD ($mm$) & ($\mu m$) & ($\mu m$) & ($\mu m$) & ($\mu m$) & \\ \hline
12.0,24.8 & 13.0,24.8 & 2.6 & 3.0 & 1.5 & 135.7 & 0.90 \\\hline
12.0,37.5 & 13.0,37.5 & 2.6 & 3.0 & 1.9 & 150.1 & 0.91 \\\hline
12.0,50.3 & 13.0,50.3 & 2.6 & 3.0 & 1.9 & 142.7 & 0.91 \\\hline
12.0,65.3 & 13.0,60.0 & 2.7 & 3.1 & 1.8 & 148.2 & 0.92 \\\hline
24.8,37.5 & 24.8,37.5 & 2.7 & 3.0 & 1.9 & 138.0 & 0.91 \\\hline
24.8,50.3 & 24.8,50.3 & 2.7 & 3.0 & 1.9 & 126.7 & 0.92 \\\hline
24.8,65.3 & 24.8,60.0 & 2.7 & 3.1 & 1.8 & 126.8 & 0.92 \\\hline
37.5,50.3 & 37.5,50.3 & 2.9 & 2.9 & 1.9 & 134.7 & 0.92 \\\hline
37.5,65.3 & 37.5,60.0 & 2.9 & 3.1 & 1.9 & 134.9 & 0.93 \\\hline
50.3,65.3 & 50.3,60.0 & 2.9 & 2.9 & 1.9 & 119.7 & 0.93 \\\hline
\end{tabular}
\end{center}
\end{table}

The traditional approach to estimating the PSF of X-ray source blur involves
assuming that there is no detector blur.
To demonstrate this approach, we estimate only the source and transmission parameters,
$s_{sx},s_{sy},l_k,$ and $h_k$ in equation \eqref{eq:blurmopt}, 
using a single set of horizontal and vertical edge radiographs
while assuming there is no detector blur, which is enforced by setting $$p^{(d)}(i,j)=\begin{cases} 1, & \text{ if } i=0,j=0\\ 0, & \text{ otherwise}\end{cases}.$$
The estimated values of X-ray source parameters for various source to object distances (SOD) 
are shown in Table \ref{tab:sourceonlypar}.
The SOD of radiographs used during 
estimation are shown in the first two columns of Table \ref{tab:sourceonlypar}.
Since FWHMs $(W_{sx},W_{sy})$ are more easily interpretable than scale parameters $(s_{sx},s_{sy})$, 
we show the FWHMs instead of the scale parameters in the third and fourth columns of Table \ref{tab:sourceonlypar}. 
We see that the estimated values for the FWHMs consistently 
increase with increasing values of SOD in Table \ref{tab:sourceonlypar}
since the source FWHMs were estimated while assuming there is no detector blur. 
Hence, the detector blur in the input radiographs get interpreted as X-ray source blur,
which causes the FWHM of X-ray source blur
to increase when SOD is increased. 

\begin{table}[h!]
\begin{center}
\caption{Mean and standard deviation of estimated parameters shown in Table \ref{tab:srcdetpar}.}
\label{tab:meanstdofpars}
\begin{tabular}{|c|c|c|c|c|c|}
\hline
 & $W_{sx}$ & $W_{sy}$ & $W_{d1}$ & $W_{d2}$ & $q$ \\\hline
\textit{Mean} & 2.7$\mu m$ & 3.0$\mu m$ & 1.8$\mu m$ & 135.7$\mu m$ & 0.92 \\\hline
\textit{Std Dev} & 0.12$\mu m$ & 0.07$\mu m$ & 0.11$\mu m$ & 9.2$\mu m$ & 0.01 \\\hline
\multirow{2}{*}{$100\frac{\textit{Std Dev}}{\textit{Mean}}$} & \multirow{2}{*}{4.4\%} & \multirow{2}{*}{2.3\%} & \multirow{2}{*}{6.1\%} & \multirow{2}{*}{6.8\%} & \multirow{2}{*}{1.09\%} \\
&&&&&\\\hline
\end{tabular}
\end{center}
\end{table}

\begin{figure}[h!]
\begin{center}
\begin{tabular}{cc}
\hspace{-0.2in}
\includegraphics[height=1.3in,keepaspectratio=true]{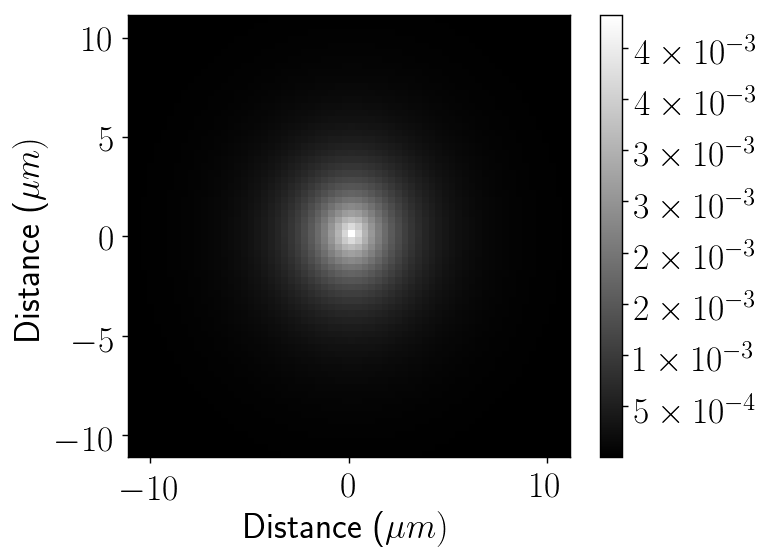} & \hspace{-0.2in} 
\includegraphics[height=1.3in,keepaspectratio=true]{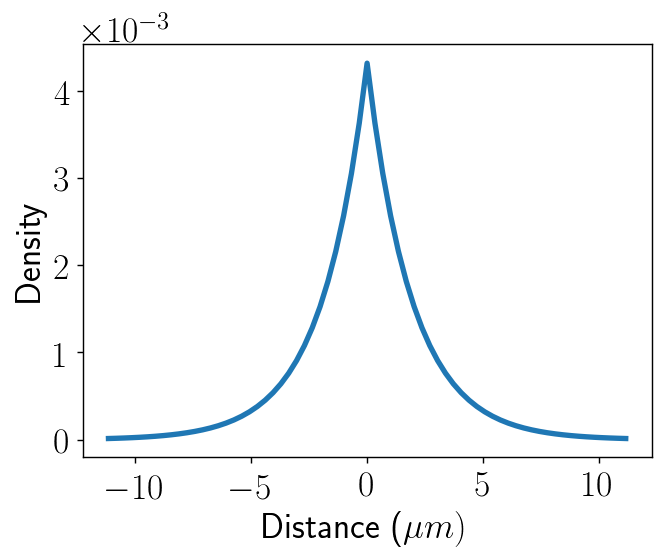} \\
\hspace{-0.2in} (a) & \hspace{-0.2in}  (b) \\
\hspace{-0.2in}
\includegraphics[height=1.3in,keepaspectratio=true]{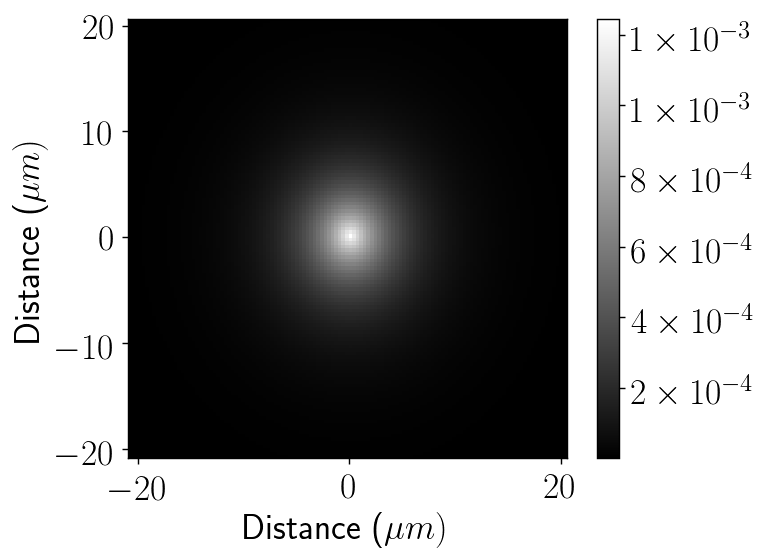} & \hspace{-0.2in}
\includegraphics[height=1.3in,keepaspectratio=true]{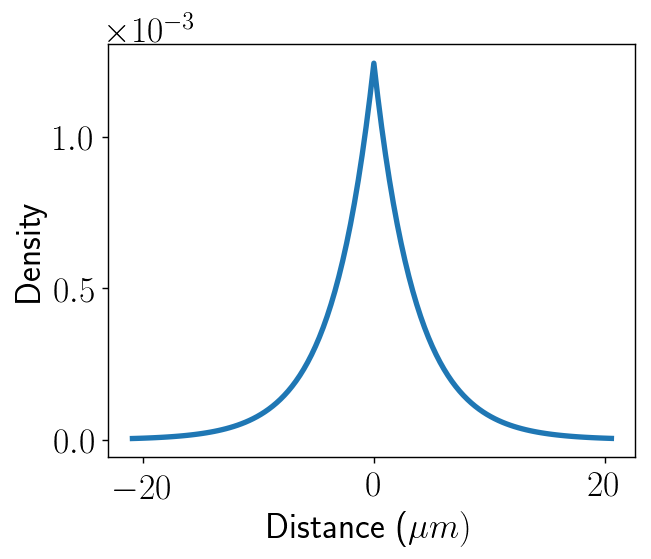} \\
\hspace{-0.2in} (c) & \hspace{-0.2in}  (d) \\
\end{tabular}
\end{center}
\caption{\label{fig:sourcepsfplot}
(a) X-ray source PSF in the source plane obtained by substituting the mean values of Table \ref{tab:meanstdofpars}
in equation \eqref{eq:srcpsf}. (c) X-ray source PSF in the detector plane 
for a SOD of $24.8mm$ and ODD of $46.2mm$ by substituting the mean values of Table \ref{tab:meanstdofpars} in equation \eqref{eq:srcdetpsf}.
(b) and (d) are line profile plots along a horizontal line through the center of the images in (a) and (c) respectively.  }
\end{figure}

\begin{figure}[!htb]
\begin{center}
\begin{tabular}{cc}
\hspace{-0.2in}
\includegraphics[height=1.4in,keepaspectratio=true]{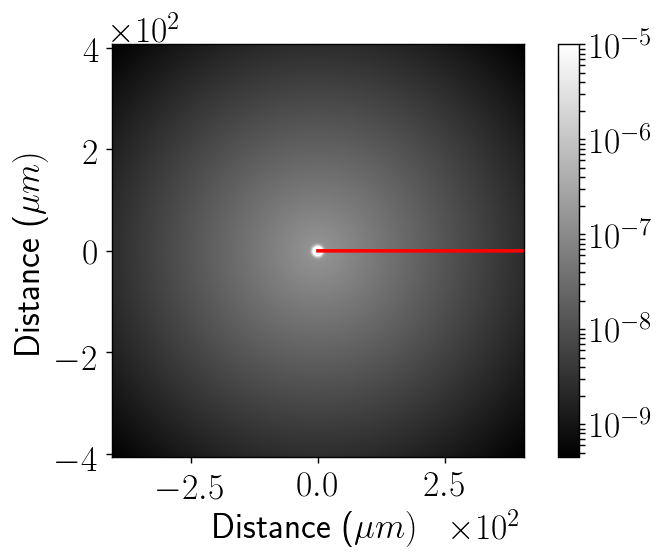} & \hspace{-0.2in}
\includegraphics[height=1.3in,keepaspectratio=true]{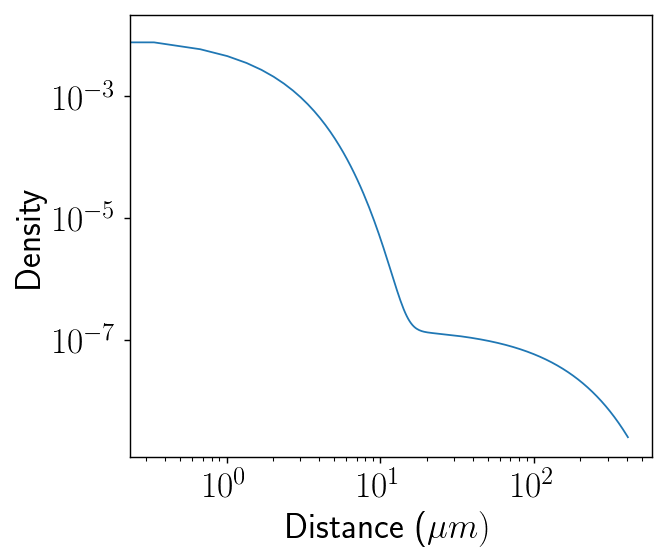}\\
\hspace{-0.2in} (a) & \hspace{-0.2in} (b) 
\end{tabular}
\end{center}
\caption{\label{fig:resdetpsfplot}
(a) Detector PSF
 in logarithmic space obtained by substituting the mean values of Table \ref{tab:meanstdofpars}
 in equation \eqref{eq:detpsf}.
 (b) Line profile plot with logarithmic axial scales
along the red colored horizontal line in (a). }
\end{figure}

The traditional approach to detector blur estimation involves assuming that there is no source blur.
To evaluate this approach, we estimate only the detector and transmission parameters,
$s_{d1},s_{d2},q,l_k,$ and $h_k$ in equation \eqref{eq:blurmopt}, 
using a single set of horizontal and vertical edge radiographs
while assuming there is no source blur, which is enforced by setting $$p^{(s)}_k(i,j)=\begin{cases} 1, & \text{ if } i=0,j=0\\ 0, & \text{ otherwise}\end{cases}.$$
The estimated parameters of detector blur 
are shown in Table \ref{tab:detonlypar}.
The SOD of radiographs used during 
estimation are shown in the first two columns of Table \ref{tab:detonlypar}.
The estimated detector blur parameters are shown in the last three columns of  Table \ref{tab:detonlypar}. 
In this case, we see that the estimated value for FWHM $W_{d1}$
decreases with increasing SOD since we assumed that there is no X-ray source blur.
Note that the first exponential with FWHM of $W_{d1}$ in equation \eqref{eq:detpsf}
is dominant since its weight given by $q$ is approximately $0.9$ in Table  \ref{tab:detonlypar}.
As SOD increases and ODD decreases, the FWHM of the source blur on the detector plane reduces.
However, since source blur is interpreted as detector blur,
the estimated FWHM $W_{d1}$ also decreases as SOD is increased.
In contrast, $W_{d2}$ and $q$ do 
not seem to have a significant dependence on SOD.

To evaluate our proposed approach, we use horizontal edge radiographs at two SODs and vertical edge radiographs at
two SODs to simultaneously estimate both source and detector blur parameters 
using algorithm \ref{alg:blurest}.
The estimated values of blur parameters after step 3 of algorithm \ref{alg:blurest} are shown in 
Table \ref{tab:srcdetpar}.
The SOD of radiographs used during 
estimation are shown in the first two columns of Table \ref{tab:srcdetpar}.
The estimated PSF parameters are shown in the last five columns of Table \ref{tab:srcdetpar}.
In this case, the estimation of X-ray source and detector
parameters are stable without any noticeable dependence on SOD.
Table \ref{tab:meanstdofpars} shows the mean (row label \textit{Mean}), standard deviation (row label \textit{Std Dev}),
and standard deviation as a percentage of the mean (row label $100\frac{\textit{Std Dev}}{Mean}$) of the estimated parameters in Table \ref{tab:srcdetpar} computed across various SODs.
From Tables \ref{tab:srcdetpar} and \ref{tab:meanstdofpars}, we can see that by simultaneously accounting for both the X-ray source
and detector blur, we are able to perform stable estimation of all PSF parameters.
 
 The source PSF as evaluated in the plane of the source and detector are shown in Fig. \ref{fig:sourcepsfplot}.
The PSF is obtained by substituting the mean values of $W_{sx}$ and $W_{sy}$ shown in Table \ref{tab:meanstdofpars}
in equations \eqref{eq:srcpsf} and \eqref{eq:srcdetpsf}.
Since $W_{sx}$ and $W_{sy}$ are approximately the same,
we can conclude that the source PSF is approximately circular in shape.
Using the method in \cite{SharmaStarPBlurM}, the manufacturer of the X-ray system estimated the source PSF to have a FWHM 
of $3.63\mu m$ using calibration data at a SOD of $10mm$ and ODD of $30mm$.
For this case, the ratio ODD/SOD lies in between the data acquisition parameters of the second and third rows in Table \ref{tab:sourceonlypar}. 
Since the manufacturer's estimate did not account for 
detector blur, their values are a good match with our estimates that also ignore detector blur. 

The detector PSF is shown in Fig. \ref{fig:resdetpsfplot}
and is evaluated by substituting the mean values of $W_{d1}$, $W_{d2}$, and $q$ shown in Table \ref{tab:meanstdofpars}
in equation \eqref{eq:detpsf}.
Since the FWHM $W_{d1}$ of the first exponential density in equation \eqref{eq:detpsf} with the high weight of $q$
is only a few pixels wide (pixel width being $0.675\mu m$), 
we can deduce that most of the detector blur is limited to only a few pixels.
Alternatively, the large value of FWHM $W_{d2}$
of the second exponential density with the smaller weight of $(1-q)$
suggests that there is significant leakage of electric charge all the way to the corners of the detector array.

\begin{table}[h!]
\begin{center}
\caption{Estimated parameters before and after step (3) of algorithm \ref{alg:blurest} using  
radiographs at SODs of $37.5mm$ and $50.3mm$.}
\label{tab:befaftstep3}
\begin{tabular}{|c|c|c|c|c|c|c|}
\hline
 & $W_{sx}$ & $W_{sy}$ & $W_{d1}$ & $W_{d2}$ & \multirow{2}{*}{$q$} \\
 & ($\mu m$) & ($\mu m$) & ($\mu m$) & ($\mu m$) & \\ \hline
Before step (3) & 4.5 & 4.6 & 2.3 & 67.6 & 0.92\\ \hline
After step (3) & 2.9 & 2.9 & 1.9 & 134.7 & 0.92\\ \hline
\end{tabular}
\end{center}
\end{table}

\begin{figure*}[!htb]
\begin{center}
\begin{tabular}{c}
\textbf{Model fit before and after step (3) of algorithm \ref{alg:blurest} for vertical edge at SOD of 37.5mm.}
\end{tabular}
\begin{tabular}{cccc}
\hspace{-0.1in}
\includegraphics[width=1.8in,height=1.8in,keepaspectratio=true]{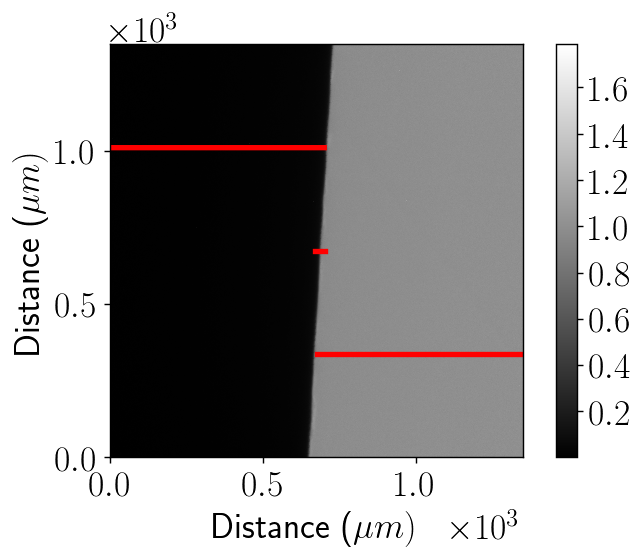} & \hspace{-0.2in}
\includegraphics[width=1.8in,height=1.8in,keepaspectratio=true]{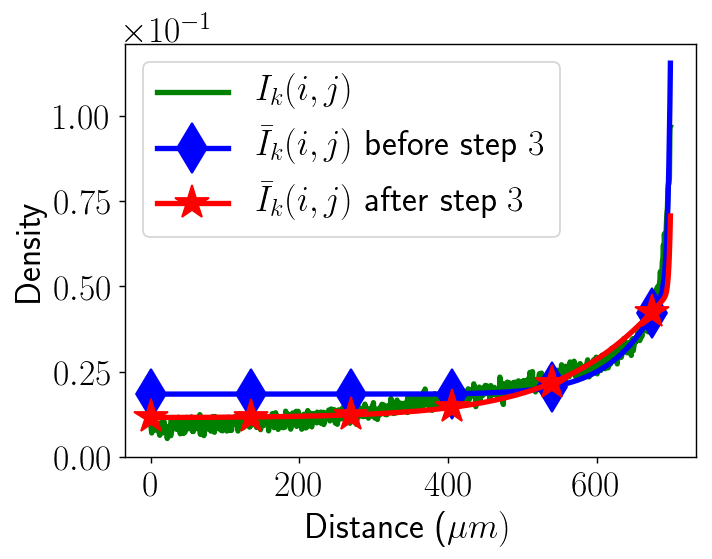} & \hspace{-0.2in}
\includegraphics[width=1.8in,height=1.8in,keepaspectratio=true]{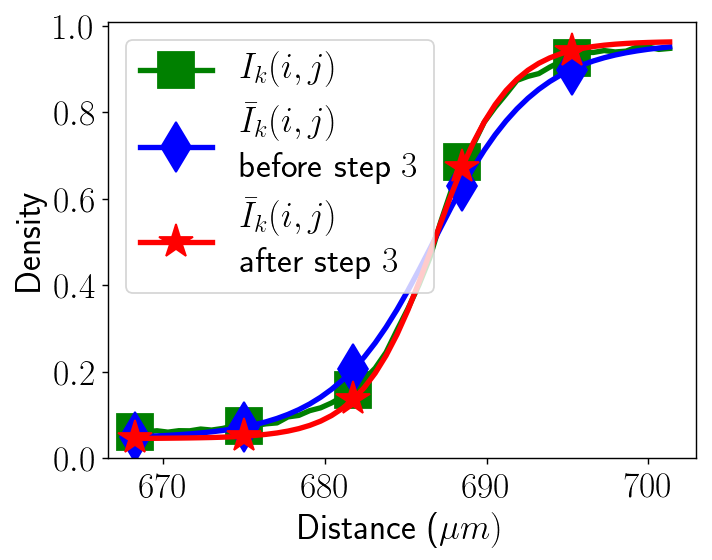} & \hspace{-0.2in}
\includegraphics[width=1.8in,height=1.8in,keepaspectratio=true]{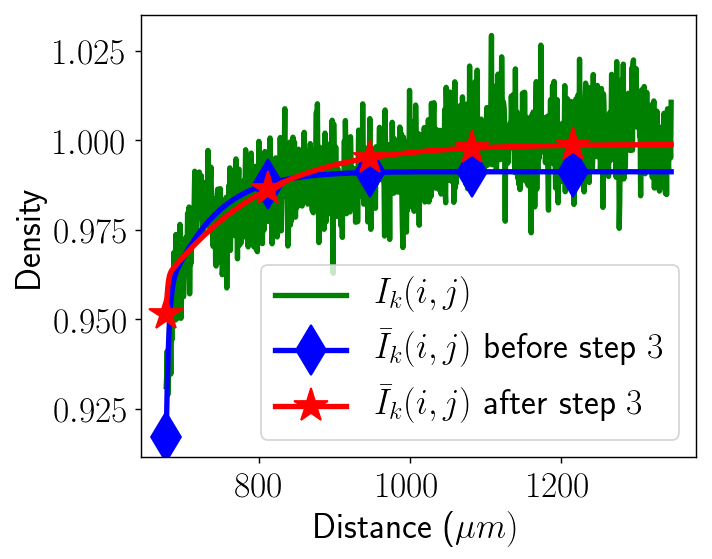}\\
\hspace{-0.1in} (a) &  \hspace{-0.1in} (b) &  \hspace{-0.2in} (c) &  \hspace{-0.2in} (d) \\
\end{tabular}
\begin{tabular}{c}
\textbf{Model fit before and after step (3) of algorithm \ref{alg:blurest} for horizontal edge at SOD of 37.5mm.}
\end{tabular}
\begin{tabular}{cccc}
\hspace{-0.1in}
\includegraphics[width=1.8in,height=1.8in,keepaspectratio=true]{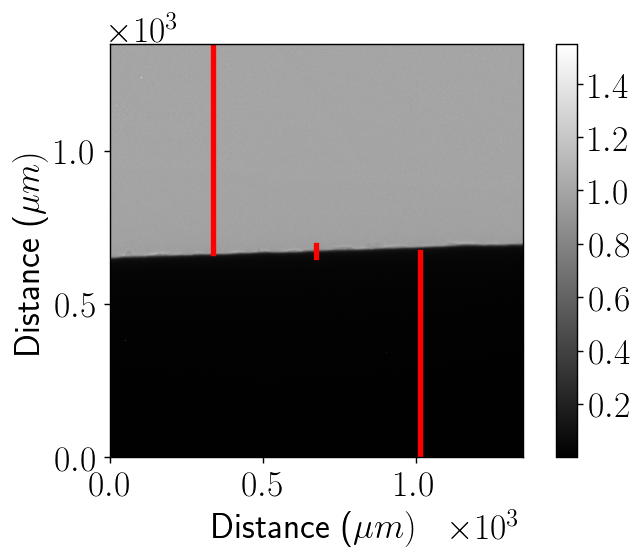} & \hspace{-0.2in}
\includegraphics[width=1.8in,height=1.8in,keepaspectratio=true]{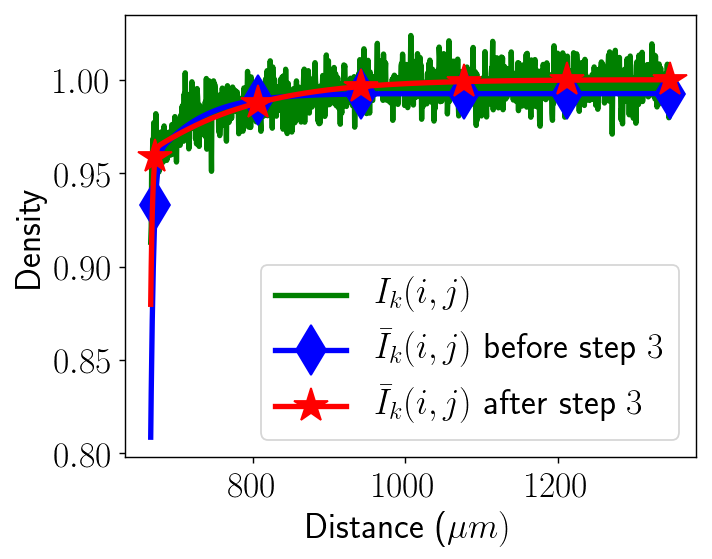} & \hspace{-0.2in}
\includegraphics[width=1.8in,height=1.8in,keepaspectratio=true]{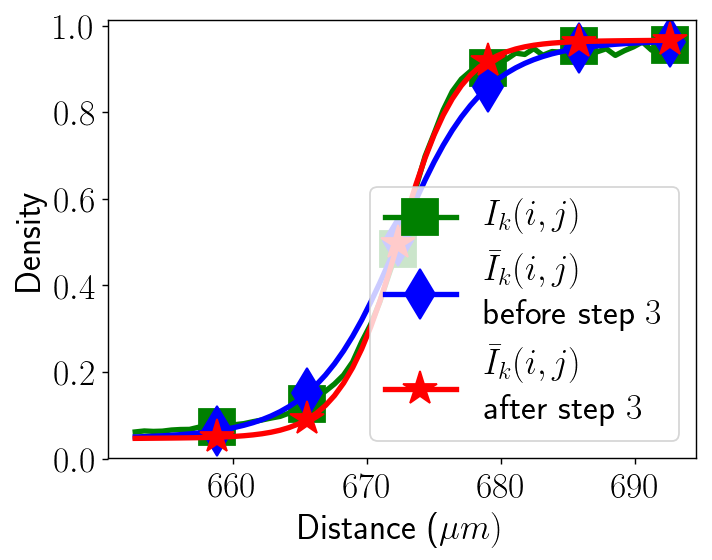} & \hspace{-0.2in}
\includegraphics[width=1.8in,height=1.8in,keepaspectratio=true]{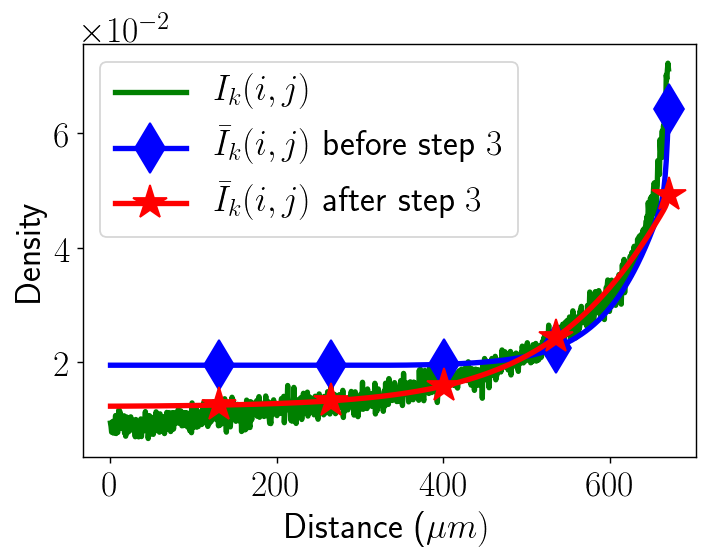} \\
\hspace{-0.1in} (e) & \hspace{-0.1in} (f) & \hspace{-0.2in} (g) & \hspace{-0.2in} (h) \\
\end{tabular}
\begin{tabular}{c}
\textbf{Fixed $q=1$ versus estimated $q$ for horizontal edge at SOD of 37.5mm.}
\end{tabular}
\begin{tabular}{cccc}
\hspace{-0.1in}
\includegraphics[width=1.8in,height=1.8in,keepaspectratio=true]{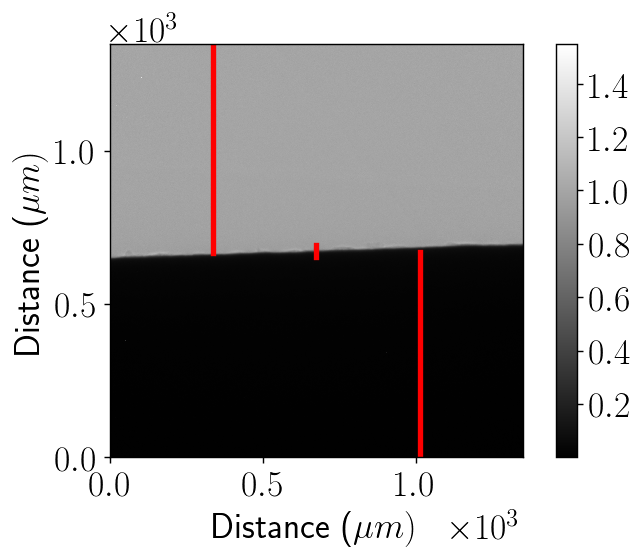} & \hspace{-0.2in}
\includegraphics[width=1.8in,height=1.8in,keepaspectratio=true]{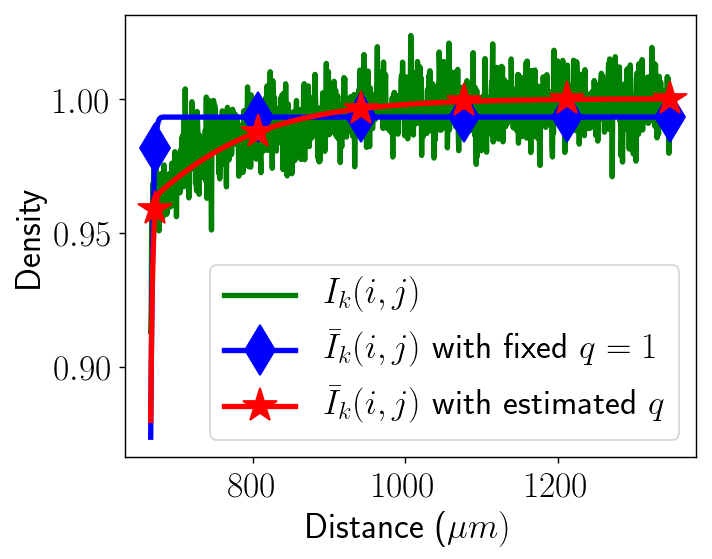} & \hspace{-0.2in}
\includegraphics[width=1.8in,height=1.8in,keepaspectratio=true]{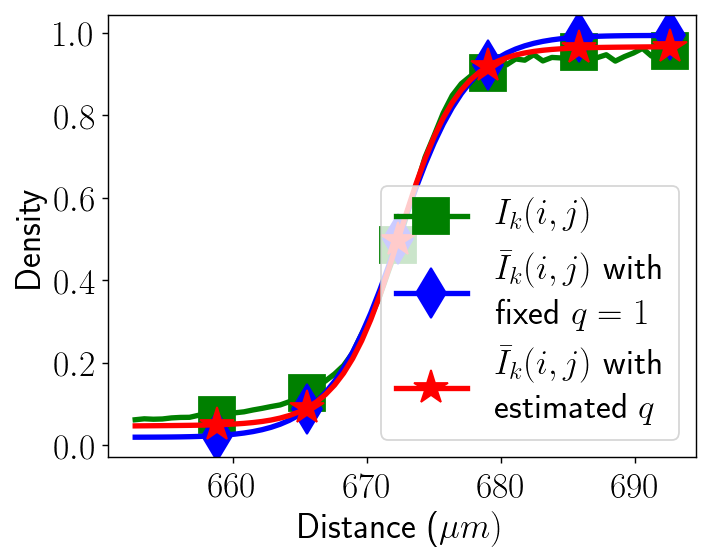} & \hspace{-0.2in}
\includegraphics[width=1.8in,height=1.8in,keepaspectratio=true]{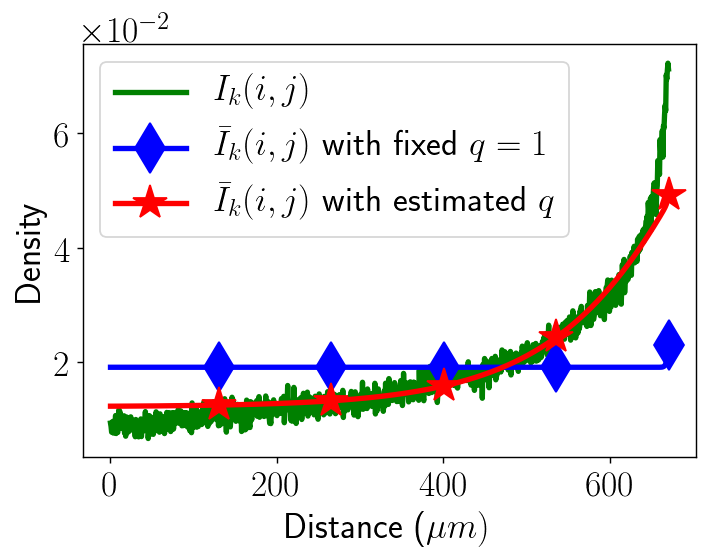} \\
\hspace{-0.1in} (i) & \hspace{-0.1in} (j)  & \hspace{-0.2in} (k) & \hspace{-0.2in} (l) \\
\end{tabular}
\end{center}
\caption{\label{fig:fitlineprof} Comparison of  blur as predicted by our model with the observed experimental blur in radiographs. Radiographs at SODs of $37.5mm$ and $50.3mm$ were used for blur estimation. 
(a,e,i) show the measured radiographs.
 (b-d), (f-h), and (j-l)
are line profile plots to evaluate the agreement between the measured radiograph $I_k(i,j)$ and the estimated radiograph using our blur model
$\bar{I}_k(i,j)=T_k (i,j) \ast p^{(s)}_{k}(i,j) \ast p^{(d)}(i,j) \ast p^{(m)}_k(i,j)$.
The plots in (b,f,j), (c,g,k), and (d,h,l) show values along the red colored lines in (a,e,i) respectively.
From (b-d) and (f-h), we can see that the accuracy of fit improves after step (3) of algorithm \ref{alg:blurest}.
From (j,l), we can see that estimating $q$ produces a better fit to the measured radiograph
in the slowly varying regions away from the sharp edge.} 
\end{figure*}

Fig. \ref{fig:fitlineprof} and Tables \ref{tab:sourceonlypar}, \ref{tab:detonlypar}, \ref{tab:srcdetpar}, and \ref{tab:befaftstep3}  validate the 
necessity to simultaneously optimize the source blur, detector blur, and transmission function. 
While step 1 and step 2 of algorithm \ref{alg:blurest} independently estimate source and detector blur,
step 3 estimates both forms of blur simultaneously. 
The estimated parameters before and after step (3) of algorithm \ref{alg:blurest} 
are shown in Table \ref{tab:befaftstep3}.
We can see that $W_{d2}$ is significantly larger after step (3) than before step (3).
This increase in $W_{d2}$ is due to the simultaneous estimation of $(l_k,h_k)$ of the transmission function
along with $s_{d2}$ of detector blur in step 3 of algorithm \ref{alg:blurest}.
To obtain a better fit, the optimizer decreases $l_k$,
increases $h_k$, which in turn allows for $W_{d2}$ to increase. 
Since $W_{d2}$ models the long tails of the detector PSF in  equation \eqref{eq:detpsf}, 
it is most sensitive to changes in $l_k$ and $h_k$.
Fig. \ref{fig:fitlineprof} (b-d,f-h) shows line profile plots that
compare the input radiograph $I_k(i,j)$ with the prediction estimate 
$\bar{I}_k(i,j)=T_k (i,j) \ast p^{(s)}_{k}(i,j) \ast p^{(d)}(i,j) \ast p^{(m)}_k(i,j)$
before and after step 3 of algorithm \ref{alg:blurest}.
The improvement in fit after step (3) as demonstrated in Fig. \ref{fig:fitlineprof} (c,g) 
is due to the change in parameters $W_{sx}$, $W_{sy}$, and $W_{d1}$.
Similarly, the improvement in fit after step (3) in Fig. \ref{fig:fitlineprof} (b,d,f,h)
is due to the change in parameters $W_{d2}$, $l_k$, and $h_k$.

Fig. \ref{fig:fitlineprof} (j-l) validates the necessity to model detector PSF as a mixture of two exponential density functions with FWHMs $W_{d1}$ and $W_{d2}$ (equation \eqref{eq:detpsf}).
The first exponential function with FWHM $W_{d1}$ and a high weight of $q$ ($q\approx 0.9$ in our experiments)
models the blur close to the sharp edge that only span a few pixels as shown in Fig. \ref{fig:fitlineprof} (k).
The second exponential function with FWHM $W_{d2}$ and a low weight of $(1-q)$ ($1-q\approx 0.1$) models the slow variation in intensity values further away from the sharp edge as shown in Fig. \ref{fig:fitlineprof} (j,l).
From Fig. \ref{fig:fitlineprof} (j,l), we can see that the detector PSF is not able to fit the slowly varying intensity that span several hundreds of pixels when $q$ is not estimated and fixed at $q=1$. Fixing $q=1$ is equivalent to modeling detector PSF as consisting of only one exponential density function with a FWHM of $W_{d1}$.

\begin{table*}[h!]
\begin{center}
\caption{Performance comparison between various choices of blur model.
Performance evaluation is on radiographs at SODs different from those used during blur estimation. 
Root mean squared error (RMSE) between the measured radiographs $I_k(i,j)$
and its predictions $\bar{I}_k(i,j)$ is used to compare blur prediction performance.
RMSE between the deblurred radiographs and the transmission functions $T_k(i,j)$
is used to compare the deblurring performance.
}
\label{tab:modelcomp}
\begin{tabular}{|c|c|c|c|c|c|c|c|c|}
\hline
& & & \multicolumn{2}{c|}{Prediction $\bar{I}_k(i,j)$} & \multicolumn{2}{c|}{Deblur, SD = $0.01$ } & \multicolumn{2}{c|}{Deblur , SD = $0.006$} \\\hline
PSF & Source & Detector & $24.8mm$/ & $65.3mm$/ & $24.8mm$/ & $65.3mm$/ & $24.8mm$/ & $65.3mm$/ \\
Model & Blur &  Blur & $24.8mm$ & $60mm$ &  $24.8mm$ & $60mm$ & $24.8mm$ & $60mm$ \\\hline
Gaussian & Yes & No & 0.0172 & 0.0157 & 0.0717 & 0.0192 & 0.0516 & 0.0172\\\hline
Exponential & Yes & No & 0.0164 & 0.0157 & 0.0387 & 0.0188 & 0.0338 & 0.0171\\\hline
Gaussian & No & Yes & 0.0228 & 0.0086 & 0.0309 & 0.0327 & 0.0296 & 0.0226\\\hline
Exponential & No & Yes & 0.0223 & 0.0083 & 0.0322 & 0.0161 & 0.0309 & 0.0132\\\hline
Gaussian & Yes & Yes & 0.0149 & 0.0081 & 0.0367 & 0.0168 & 0.0289 & 0.0126\\\hline
Exponential & Yes & Yes & \textbf{0.0147} & \textbf{0.0078} & \textbf{0.0264} & \textbf{0.0139} & \textbf{0.0243} & \textbf{0.0111}\\\hline
\end{tabular}
\end{center}
\end{table*}

\begin{figure}[!htb]
\begin{center}
\begin{tabular}{cc}
\hspace{-0.1in}
 \includegraphics[width=1.5in,height=1.5in,keepaspectratio=true]{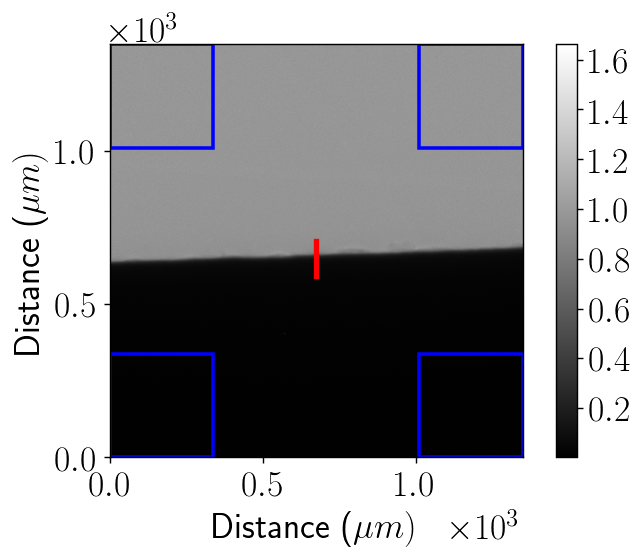} & \hspace{-0.2in}
\includegraphics[width=1.5in,height=1.5in,keepaspectratio=true]{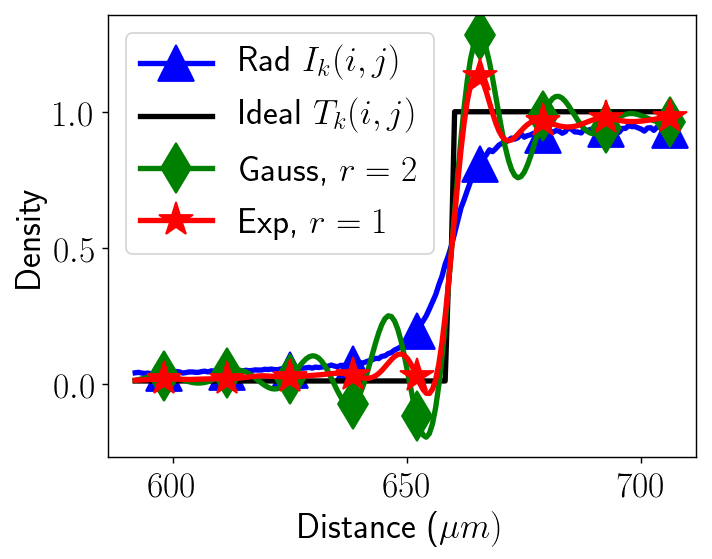} \\
 \hspace{-0.1in} (a) SD = $0.01$ &  \hspace{-0.1in} (b) SD = $0.01$ \\
  \includegraphics[width=1.5in,height=1.5in,keepaspectratio=true]{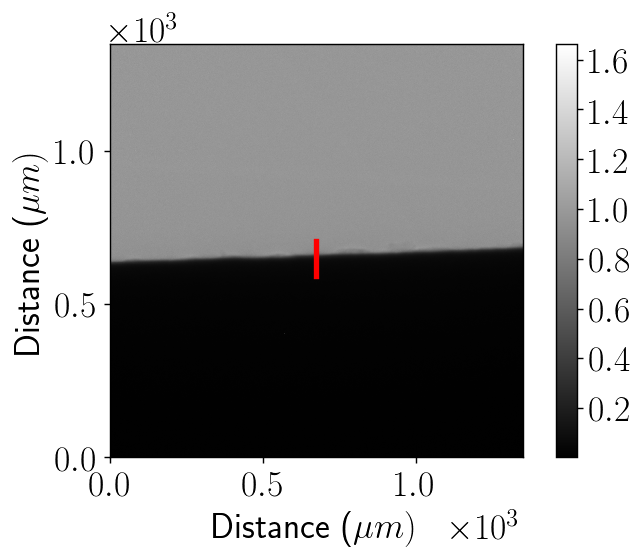} & \hspace{-0.2in}
\includegraphics[width=1.5in,height=1.5in,keepaspectratio=true]{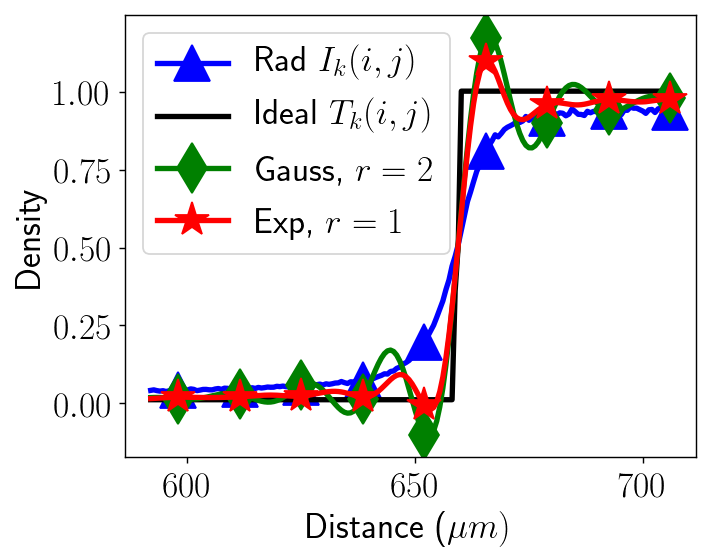} \\
 \hspace{-0.1in} (c) SD = $0.006$ &  \hspace{-0.1in} (d) SD = $0.006$ \\
\end{tabular}
\end{center}
\caption{\label{fig:debedgeprof}
Comparison of deblurring performance when both X-ray source and detector PSFs are modeled as either exponential ($r=1$) or Gaussian ($r=2$).
(b,d) are line profiles along the red line in the deblurred radiographs (a,c) respectively.
The $1^{st}$ and $2^{nd}$ rows 
show the deblured result for noise levels of SD = $0.01$ and SD = $0.006$ respectively.
To compute the noise level, the standard deviations within square rectangular boxes (positions marked as blue boxes in (a)) 
at the four corners of the radiograph are computed and averaged.
In (b,d), the severity of ringing artifacts is worse with the Gaussian model when compared to exponential model.
}
\end{figure}

\begin{figure}[!htb]
\begin{center}
\includegraphics[width=2.8in,keepaspectratio=true]{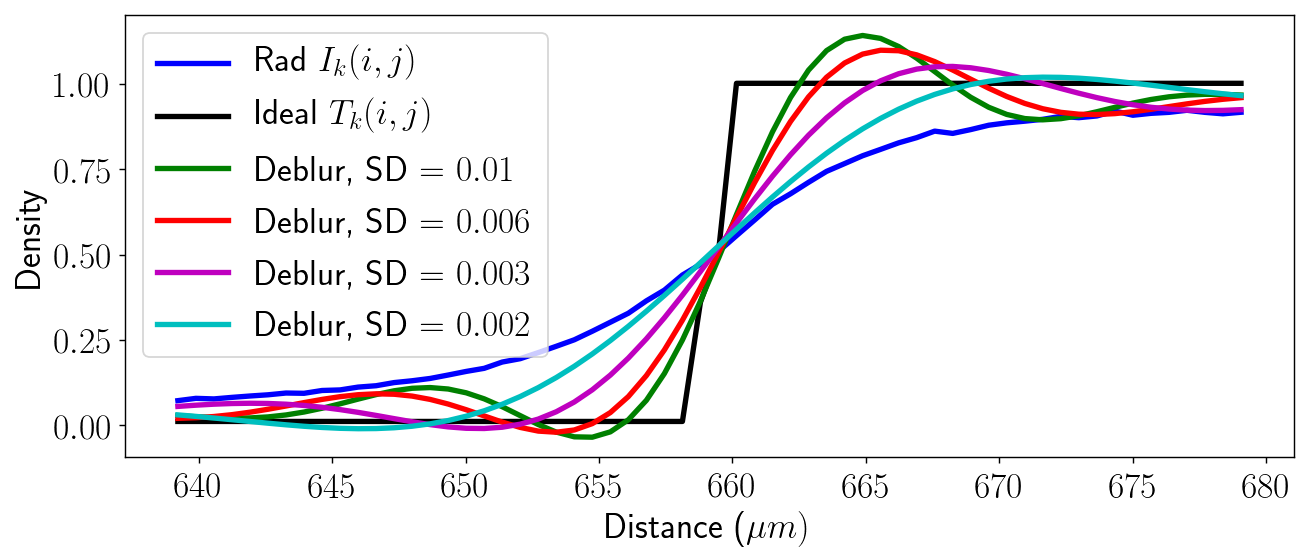} \\
\end{center}
\caption{\label{fig:debedgereg}
Comparison of deblurred edge at various noise levels obtained using different regularization parameters.
Both source and detector PSFs are modeled using exponential density function ($r=1$).
When noise level in deblurred radiograph is decreased by increasing regularization, 
both the ringing artifacts and the sharpness also reduce. 
}
\end{figure}

\begin{table}[h!]
\begin{center}
\caption{Variation in deblurring performance 
due to SOD dependent variation in estimated PSF parameters (Table \ref{tab:srcdetpar}).
The last two columns show the RMSE between the transmission functions $T_k(i,j)$
and the deblurred radiographs at SOD of $24.8mm$.
}
\label{tab:flucdeb}
\begin{tabular}{|c|c|c|c|}
\hline
Horizontal & Vertical & Deblur & Deblur \\
SOD ($mm$) & SOD ($mm$) & SD = $0.01$ & SD = $0.006$ \\\hline
12.0,37.5 & 13.0,37.5 & 0.0242 & 0.0222\\\hline
12.0,50.3 & 13.0,50.3 & 0.0246 & 0.0226\\\hline
12.0,65.3 & 13.0,60.0 & 0.0241 & 0.0220\\\hline
37.5,50.3 & 37.5,50.3 & 0.0264 & 0.0243\\\hline
37.5,65.3 & 37.5,60.0 & 0.0258 & 0.0236\\\hline
50.3,65.3 & 50.3,60.0 & 0.0262 & 0.0242\\\hline
\end{tabular}
\vspace{0.2in}
\caption{Quantifying the varying RMSE performance along the last two columns of Table \ref{tab:flucdeb}.}
\label{tab:flucdebstat}
\begin{tabular}{|c|c|c|}
\hline
 & Deblur & Deblur \\
 & SD = $0.01$ & SD = $0.006$ \\\hline
Mean & 0.0241 & 0.0220\\\hline
Std Dev & 0.0010 & 0.0009\\\hline
\multirow{2}{*}{$100\frac{\text{Std Dev}}{\text{Mean}}$} & \multirow{2}{*}{4.15\%} & \multirow{2}{*}{4.09\%} \\
&&\\\hline
\end{tabular}
\end{center}
\end{table}

\subsection{Validating the Blur Model}
The optimal choice of density function used to model the source blur in equation \eqref{eq:srcpsf} and the detector blur in equation \eqref{eq:detpsf} is system dependent. 
In Table \ref{tab:modelcomp}, we compare the performance of various choices of blur model
using quantitative evaluation metrics.
Horizontal and vertical edge radiographs at SODs of $37.5mm$ and $50.3mm$ were used when simultaneously estimating both source and detector blurs.
When estimating only source blur, only radiographs at SOD of $37.5mm$ were used since source blur is dominant at the lower SOD. 
When estimating only detector blur, only radiographs at SOD of $50.3mm$ were used since detector blur is dominant at the higher SOD. 
The $1^{st}$ column of Table \ref{tab:modelcomp} indicates whether the 
PSF model was chosen to be either Gaussian ($r=2$) or exponential ($r=1$).
The $2^{nd}$ column indicates if source blur was modeled (marked as ``Yes") or ignored (marked as ``No").
The $3^{rd}$ column indicates if detector blur was modeled (marked as ``Yes") or ignored (marked as ``No").
To determine the most appropriate blur model, we evaluate the blur prediction 
and deblurring performance on radiographs that were acquired at SOD values 
different than those used for blur estimation.
Computation of the performance evaluation metric, root mean squared error (RMSE), in the $4^{th},6^{th}$, and $8^{th}$ columns is over both the horizontal and vertical edge radiographs at SOD of $24.8mm$.
Similarly, the RMSEs in the $5^{th},7^{th}$, and $9^{th}$ columns are computed over the horizontal edge radiograph at SOD of $65.3mm$ and vertical edge radiograph at SOD of $60mm$.  

The accuracy of blur prediction is quantified in the $4^{th}$ and $5^{th}$ columns 
under the header ``Prediction $\bar{I}_k(i,j)$" 
by computing the RMSE between the measured radiographs $I_k(i,j)$ and its predictions $\bar{I}_k(i,j)=T_k (i,j) \ast p^{(s)}_{k}(i,j) \ast p^{(d)}(i,j) \ast p^{(m)}_k(i,j)$ using the blur model.
In order to compute $\bar{I}_k(i,j)$, we must compute the transmission function $T_k(i,j)$
by setting appropriate values for $l_k,h_k$ in equation \eqref{eq:transvsideal} for radiographs that were not used during blur estimation.
As an approximate solution to this problem, the $l_k$ and $h_k$ for horizontal/vertical edge prediction $\bar{I}_k(i,j)$ are chosen to be the average of the corresponding values for the horizontal/vertical edge transmission functions estimated during blur optimization.
The lowest RMSE is obtained along the last row of Table \ref{tab:modelcomp}, 
which corresponds to simultaneous estimation of both source and detector blurs 
while assuming an exponential density function for both PSFs. 

 \begin{figure*}[!htb]
\begin{center}
\begin{tabular}{cccc}
\hspace{-0.3in}
 \includegraphics[width=1.65in,keepaspectratio=true]{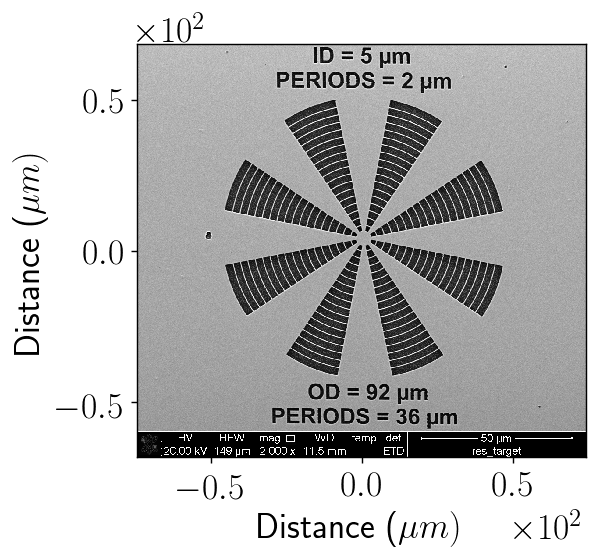}  & \hspace{-0.2in}
 \includegraphics[width=1.45in,keepaspectratio=true]{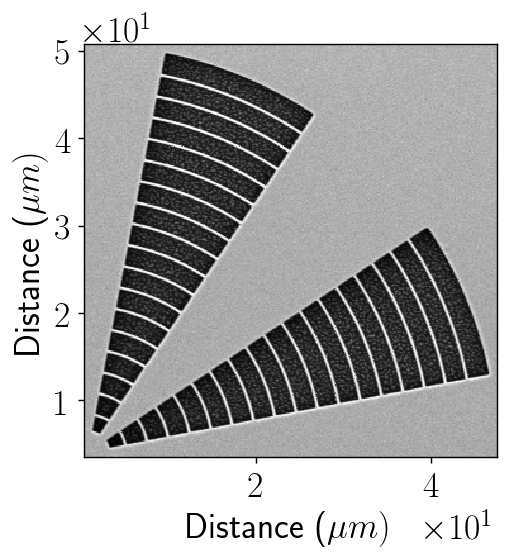} & \hspace{-0.2in}
\includegraphics[width=1.9in,keepaspectratio=true]{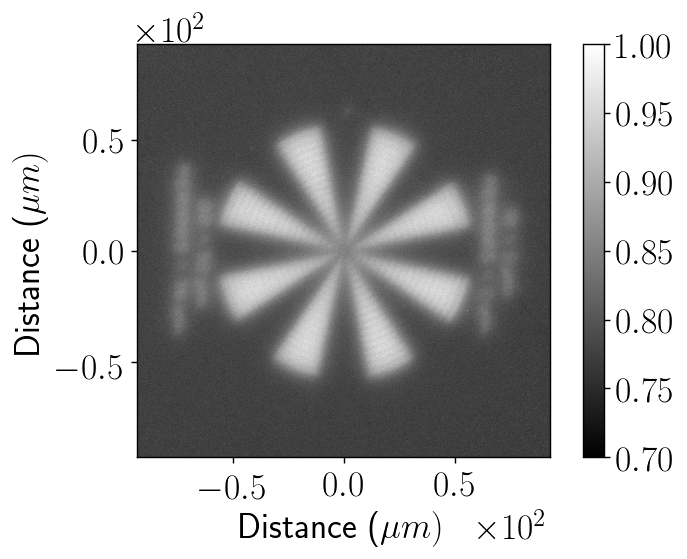} & \hspace{-0.2in}
\includegraphics[width=1.8in,keepaspectratio=true]{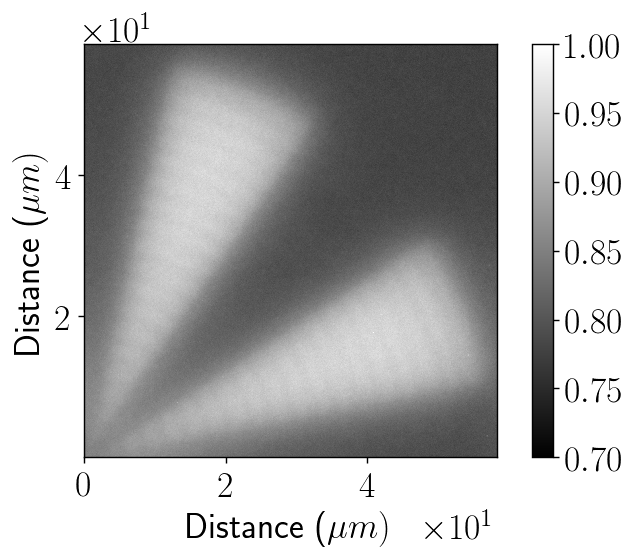} \\
\hspace{-0.2in} (a) Artifact & \hspace{-0.2in} (b) Artifact zoom & \hspace{-0.2in} (c) Measured & \hspace{-0.2in} (d) Measured zoom\\
\hspace{-0.2in}
\includegraphics[width=1.9in,keepaspectratio=true]{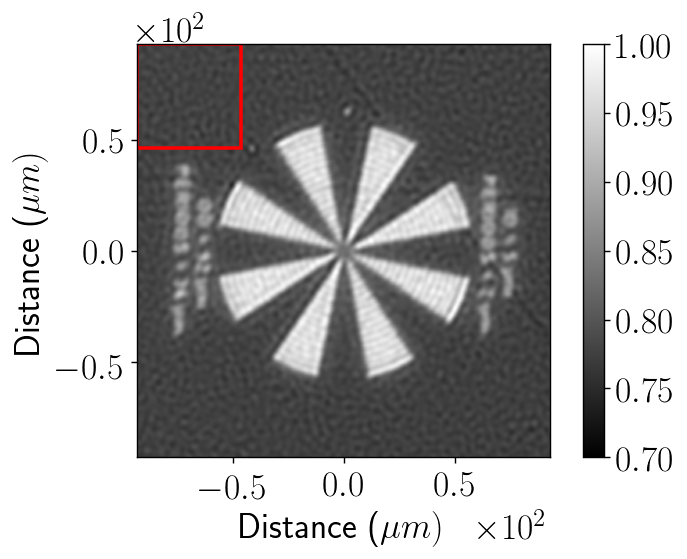} & \hspace{-0.2in}
\includegraphics[width=1.8in,keepaspectratio=true]{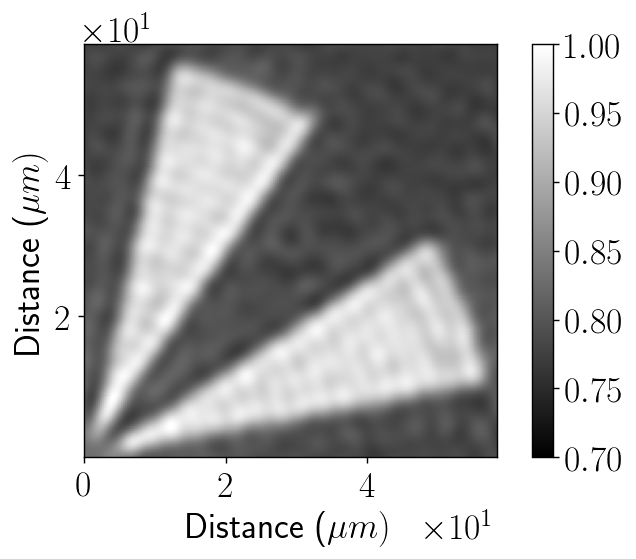} & \hspace{-0.2in}
\includegraphics[width=1.9in,keepaspectratio=true]{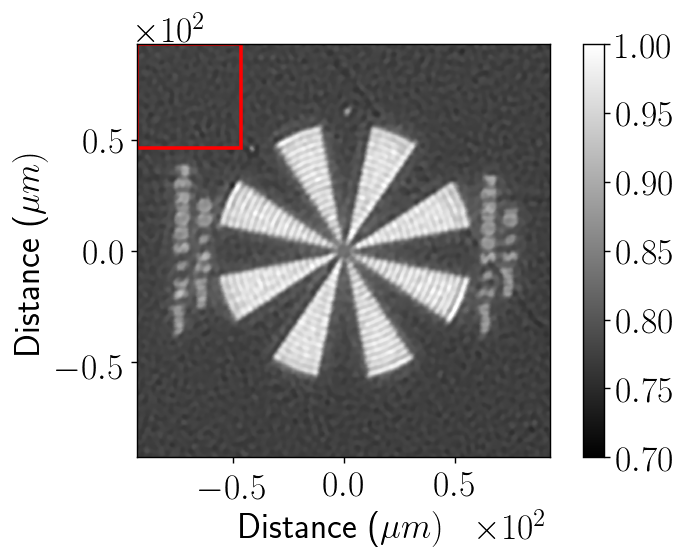} & \hspace{-0.2in}
\includegraphics[width=1.8in,keepaspectratio=true]{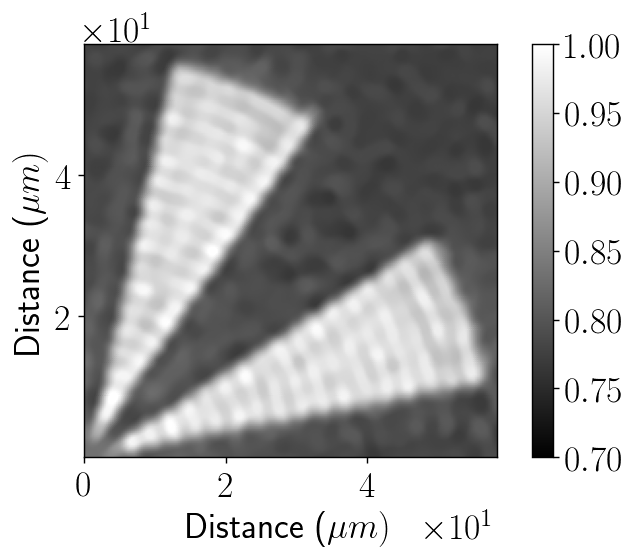} \\
\hspace{-0.2in} (e) Wiener  & \hspace{-0.2in} (f) Wiener zoom & \hspace{-0.2in} (g) RLSD  & \hspace{-0.2in} (h) RLSD zoom\\
\end{tabular}
\end{center}
\caption{\label{fig:deblur} 
Deblurring radiographs of a star shaped pattern. (a) shows a star shaped artifact imaged using scanning electron microscopy (SEM).
(c) shows the X-ray radiograph of the artifact. (e) shows the X-ray radiograph that is deblurred 
using Wiener filtering. (g) shows the X-ray radiograph that is deblurred
using RLSD.
(b), (d), (f), and (h) zooms into the top right corner of the images in 
(a), (c), (e), and (g) respectively. Both Wiener filtering and RLSD reduce blur.
RLSD better preserves sharpness when compared to Wiener filtering. }
\end{figure*}

Next, we compare the deblurring performance using Wiener filtering for the different choices of blur model.
Since Wiener filter uses a regularization parameter that trades off noise with resolution,
it is not meaningful to compare the quality of the edge when the deblurred radiographs under comparison have varying noise levels.  
Hence, the regularization parameter for Wiener filter is adjusted until the noise level in the radiograph is close to a pre-specified constant $\gamma$.
Fig. \ref{fig:debedgeprof} and $6^{th}$ to $9^{th}$ columns under the headers ``Deblur, SD = $\gamma$" in Table \ref{tab:modelcomp} compare the deblurring performance for various model choices.
In Fig. \ref{fig:debedgeprof}, deblurred results are shown for horizontal edge radiograph at a SOD of $24.8mm$
while blur estimation was performed using radiographs at SODs of $37.5mm$ and $50.3mm$. 
The noise level $\gamma$ is computed as the average of
the standard deviations within rectangular boxes along each of the four corners of a radiograph (similar to the blue squares in Fig. \ref{fig:debedgeprof} (a)).
The line profiles in Fig. \ref{fig:debedgeprof} (b,d) are along the red line in  Fig. \ref{fig:debedgeprof} (a,c) respectively. 
The RMSE for deblurring comparison in Table \ref{tab:modelcomp} 
is computed between the deblurred radiographs and its corresponding transmission functions given by equation $\eqref{eq:transvsideal}$. 
The lowest RMSE under the header ``Deblur, SD = $\gamma$" is achieved along the last row of Table  \ref{tab:modelcomp}.
In Fig. \ref{fig:debedgeprof}, we can see that 
using exponential over Gaussian density function results in less intense ringing artifacts in (b,d).
Thus, the best choice of blur model is to use exponential density function ($r=1$)
over Gaussian density function ($r=2$) while simultaneously modeling the effects
of both X-ray source and detector blur.
Fig. \ref{fig:debedgereg} shows that reducing noise and ringing artifacts in deblurred radiograph
using regularization also reduces the sharpness of the Tungsten edge.

Table \ref{tab:flucdeb} compares the variation in deblurring performance as a consequence of the SOD dependent variation in estimated PSF parameters in Table \ref{tab:srcdetpar}.
Both source and detector blur PSFs were modeled using exponential density function ($r=1$)
and estimated using algorithm \ref{alg:blurest}.
From Table \ref{tab:flucdeb}, we can see that irrespective of the SOD used to estimate blur, the RMSE is close to the corresponding value in the last row of Table \ref{tab:modelcomp} (under column header ``Deblur, SD = $\gamma$" and sub-header ``24.8mm/24.8mm").
Remarkably, this RMSE for a joint source and detector blur model with exponential density is always lower than the RMSE for all other model choices considered in Table \ref{tab:modelcomp}.  
Table \ref{tab:flucdebstat} computes the mean, standard deviation, and standard deviation as a percentage of the mean for each column in Table \ref{tab:flucdeb}.
The standard deviation as a percentage of the mean shown in the last row of Table \ref{tab:flucdebstat} is in the same general range as the corresponding value computed for the PSF parameters in the last row of Table \ref{tab:meanstdofpars}.

\subsection{Deblurring Algorithms}
A variety of deblurring algorithms that use the  estimated X-ray source and detector PSFs can be used to deblur radiographs.  
We acquired radiograph (Fig. \ref{fig:deblur} (c,d)) of a star shaped object at
a SOD of $10mm$ and ODD of $61mm$.
This star shaped test object 
was composed of a $1\mu m$ thick Tungsten layer on a SiN membrane.
The blurry radiograph in Fig. \ref{fig:deblur} (c) is then deblurred using Wiener filtering
and RLSD algorithm.
The regularization of both Wiener filter and RLSD methods
are adjusted until the noise variance in the red square region 
in Fig. \ref{fig:deblur} (e) and (g) are the same and equal to $0.005$.
The deblurred radiographs shown in Fig. \ref{fig:deblur} (e-h)
are much sharper than the input radiograph in Fig. \ref{fig:deblur} (c,d). 
A scanning electron microscopy (SEM) image of the star artifact in Fig. \ref{fig:deblur} (a,b)
show the slits inside each spoke of the star pattern. 
By comparing Fig. \ref{fig:deblur} (f) with 
Fig. \ref{fig:deblur} (h), we see that the sharpness of RLSD image is better than the Wiener image
since the slits in the spokes of the star pattern are more clear in the RLSD image.

\section{Conclusion}
In this paper, we presented a method to estimate both X-ray source and detector blur
from radiographs of a Tungsten plate rollbar. Importantly, our method is able to distentangle 
and estimate the parameters of both X-ray source and detector blur from radiographs 
that are simultaneously blurred by both forms of blur.
We show that blur estimation can be performed using horizontal edge and vertical edge radiographs,
each of which is measured at two different values of the ratio of object to detector distance (ODD) and source to object distance (SOD).
Using the estimated blur model, we demonstrated the ability to deblur radiographs using various deblurring algorithms.


%

\appendices
\section{Computing the Ideal Transmission Function}
\label{app:itransfunc}
For every radiograph $I_k(i,j)$, the ideal transmission function $\tilde{T}_k(i,j)$ is estimated from $I_k(i,j)$ using
traditional image processing algorithms. 
The first step to computing $\tilde{T}_k(i,j)$ is to determine the location of the Tungsten plate's boundary.
Even with the rolled edge, the plate is designed such that the transition
region from Tungsten to air is no more than one pixel thick.
But, from Fig. \ref{fig:blurvsidealrad} (a), we can see that it is very difficult to 
determine the exact pixel locations of the plate's boundary due to blur from the X-ray source and detector.
Hence, we scale the radiograph $I_k(i,j)$ such that it is in the range of $0$ to $1$ and assume that the edge  
lies along a iso-valued contour with a level value of $0.5$.
Such a iso-valued contour is estimated using the marching squares algorithm \cite{marchcubes} implemented in the python package
\textit{scikit-image} \cite{scikit-image}.

The ideal transmission function $\tilde{T}_k(i,j)$
is assigned a value of $0$ for the pixels belonging to the Tungsten plate and
$1$ for the pixels with no plate (or air pixels) since the plate is designed to completely attenuate all X-rays.
For the edge boundary pixels, their values are linearly interpolated given a value of $0.5$ along the iso-valued contour and neighboring pixel values of 1 and 0.
The iso-valued contour produced by the marching squares algorithm is in the form of a list of real valued $(k,l)$ coordinates. However, the pixel coordinates $(i,j)$ of $\tilde{T}_k(i,j)$ are integer valued. Hence, the values at edge pixel coordinates $(\lfloor k\rfloor,\lfloor l\rfloor)$, $(\lfloor k\rfloor,\lceil l\rceil)$, $(\lceil k\rceil,\lfloor l\rfloor)$, and $(\lceil k\rceil,\lceil l\rceil)$ are determined by interpolation.

To prevent aliasing during convolution, $\tilde{T}_k(i,j)$ is padded to three times the size of $I_k(i,j)$ by a special padding procedure that takes into 
account the orientation of the Tungsten plate (Fig. \ref{fig:blurvsidealrad} (b)).
The same amount of padding is applied to the top, bottom, left, and right
edge of the images. 
First, a straight line is fit to the sharp edge of the Tungsten plate (Fig. \ref{fig:blurvsidealrad} (a)). 
We extend the straight lines outside the image
to account for the plate extending outside the image's field of view.
The straight line will split the padded image into two regions, 
one with the Tungsten plate and another without the plate. 
Every padded pixel will have a value of $0$ or $1$ depending on
whether it lies in the region with the plate or without the plate.

\section{Gradient Computation}
\label{app:objgrad}
To solve the optimization problem in equations \eqref{eq:algopt1}, \eqref{eq:algopt2}, and \eqref{eq:algopt3},
we use the L-BFGS-B algorithm \cite{LBFGSBTheory}. Similar to many optimization algorithms such as
gradient descent and conjugate gradient, L-BFGS-B needs a routine to calculate
the gradient of the objective function in equations \eqref{eq:algopt1}, \eqref{eq:algopt2}, and \eqref{eq:algopt3}
with respect to the variables that are being optimized.
For example, to solve the optimization problem in  \eqref{eq:algopt1},
L-BFGS-B needs to know the gradient of the objective function $\sum_{k\in \Omega_s} E_k$
with respect to $s_{sx}$ and $s_{sy}$.

We will derive the gradient with respect to all variables for the objective function in \eqref{eq:algopt3} (same as \eqref{eq:blurmopt}).
Gradients for solving \eqref{eq:algopt1} and \eqref{eq:algopt2} will be a straightforward extension of the gradient
derived for \eqref{eq:algopt3}.
Let the objective function be denoted by $f$ i.e, $f = \sum_{k=1}^K E_k$.
To solve \eqref{eq:algopt3}, we need the gradient of $f$, which is a vector 
consisting of the partial derivatives $\frac{\partial f}{\partial s_{sx}}$, $\frac{\partial f}{\partial s_{sy}}$, 
$\frac{\partial f}{\partial s_{d1}}$, $\frac{\partial f}{\partial s_{d2}}$,
$\frac{\partial f}{\partial q}$, $\frac{\partial f}{\partial l_k}$, and $\frac{\partial f}{\partial h_k}\,\, \forall k$.

To compute the partial derivatives, we use the chain rule and quotient rule of calculus.
First, we will partially compute only that part of the derivative which is common 
to all parameters irrespective of whether it belongs to the source PSF, the detector PSF, or the transmission function.
Let $v$ represent any one parameter among $s_{sx}$, $s_{sy}$, $s_{d1}$, $s_{d2}$, $q$,
$l_k$, and $h_k$. 
Then, the derivative of $f$ with respect to $v$ is,
\begin{equation}
\label{eq:derivwrtv}
\frac{\partial f}{\partial v} = \sum_k\sum_{i,j} w_k(i,j)\left(I_k(i,j)-\bar{I}_k(i,j)\right) \frac{\partial \bar{I}_k(i,j)}{\partial v}
\end{equation}
where $\bar{I}_k(i,j)=T_k (i,j) \ast p^{(s)}_{k}(i,j) \ast p^{(d)}(i,j) \ast p^{(m)}_k(i,j)$.
In equation \eqref{eq:derivwrtv}, every term and operator except $\frac{\partial \bar{I}_k(i,j)}{\partial v}$
is independent of whether $v$ is $s_{sx}$, $s_{sy}$, $s_{d1}$, $s_{d2}$, $q$,
$l_k$, or $h_k$. 
Next, we shall expand the derivative $\frac{\partial \bar{I}_k(i,j)}{\partial v}$.
This requires us to consider source, detector, and transmission function parameters separately.

\subsection{Variable $v$ is a source PSF parameter}
Let $v$ be one among the source PSF parameters $s_{sx}$ or $s_{sy}$.
For compactness of representation, 
we shall assume $g_{s,k}(i,j)=\exp\left( -\Delta^r \frac{SOD^r_k}{ODD^r_k}\left(i^2 s^2_{sx} +j^2 s^2_{sy} \right)^{\frac{r}{2}} \right)$
and $Z_{s,k} = \sum_{i,j}g_{s,k}(i,j)$ from equation \eqref{eq:srcdetpsfnorm}. Then, using the quotient rule,
\begin{multline}
\frac{\partial \bar{I}_k(i,j)}{\partial v} = T_k (i,j) \ast 
\frac{1}{Z_{s,k}^2}\left(Z_{s,k}\frac{\partial g_{s,k}(i,j)}{\partial v} - \right. \\
\left. g_{s,k}(i,j)\frac{\partial Z_{s,k}}{\partial v}\right) 
\ast p^{(d)}(i,j) \ast p^{(m)}_k(i,j),
\end{multline}
where $\ast$ denotes discrete 2D convolution,
$\partial Z_{s,k}/\partial v = \sum_{i,j} \partial g_{s,k}(i,j)/\partial v$, and
\begin{equation}
\frac{\partial g_{s,k}(i,j)}{\partial v} = \begin{cases}
	  -r g_{s,k}(i,j) \Delta^r \frac{SOD^r_k}{ODD^r_k} s_{sx} \left(s^2_{sx} i^2+s^2_{sy} j^2\right)^{\frac{r}{2}-1} i^2,\\ \quad\quad\quad\quad\quad\quad\quad\quad\quad\quad\quad\quad\quad\quad\text{ if $v$ is $s_{sx}$},\\
	  -r g_{s,k}(i,j) \Delta^r \frac{SOD^r_k}{ODD^r_k} s_{sy} \left(s^2_{sx} i^2+s^2_{sy} j^2\right)^{\frac{r}{2}-1} j^2,\\ \quad\quad\quad\quad\quad\quad\quad\quad\quad\quad\quad\quad\quad\quad\text{ if $v$ is $s_{sy}$}.\\
\end{cases}
\end{equation}

\subsection{Variable $v$ is a detector PSF parameter}
Let $v$ be one among the detector PSF parameters $s_{d1}$, $s_{d2}$, or $q$.
For compactness of representation, 
we shall assume $g_{d1}(i,j)=\exp\left(-s^r_{d1} \Delta^r \left(i^2+j^2\right)^{\frac{r}{2}}\right)$, $g_{d2}(i,j)=\exp\left(-s^r_{d2} \Delta^r \left(i^2+j^2\right)^{\frac{r}{2}} \right)$,
 $Z_{d1} = \sum_{i,j}g_{d1}(i,j)$, and $Z_{d2} = \sum_{i,j}g_{d2}(i,j)$ from equations \eqref{eq:detpsfnorm1} and \eqref{eq:detpsfnorm2}. 
Using the quotient rule, we get,
\begin{multline}
\hspace{-0.2in}\frac{\partial \bar{I}_k(i,j)}{\partial v} =  T_k (i,j) \ast p^{(s)}_k(i,j) \ast \frac{\partial p^{(d)}(i,j)}{\partial v} \ast p^{(m)}_k(i,j),
\end{multline}
where $\ast$ denotes discrete 2D convolution,
\begin{equation*}
\frac{\partial p^{(d)}(i,j)}{\partial v} = \begin{cases}
	\frac{q}{Z_{d1}^2}\left(Z_{d1}\frac{\partial g_{d1}(i,j)}{\partial v} - g_{d1}(i,j)\frac{\partial Z_{d1}}{\partial v}\right), & \text{if $v$ is $s_{d1}$},\\
	\frac{1-q}{Z_{d2}^2}\left(Z_{d2}\frac{\partial g_{d2}(i,j)}{\partial v} - g_{d2}(i,j)\frac{\partial Z_{d2}}{\partial v}\right), & \text{if $v$ is $s_{d2}$},\\
	\frac{1}{Z_{d1}}g_{d1}(i,j) - \frac{1}{Z_{d2}}g_{d2}(i,j), & \text{if $v$ is $q$},
\end{cases}
\end{equation*}
$\partial g_{d1}(i,j)/\partial s_{d1}=-g_{d1}(i,j) \Delta^r \left(i^2+j^2\right)^{\frac{r}{2}}r s_{d1}^{r-1}$, $\partial g_{d2}(i,j)/\partial s_{d2}=-g_{d2}(i,j) \Delta^r \left(i^2+j^2\right)^{\frac{r}{2}}r s_{d2}^{r-1}$,
$\partial Z_{d1}/\partial v = \sum_{i,j} \partial g_{d1}(i,j)/\partial v$, and 
$\partial Z_{d2}/\partial v = \sum_{i,j} \partial g_{d2}(i,j)/\partial v$.
 
\subsection{Variable $v$ is a transmission function parameter}
Let $v$ be one among the transmission function parameters $l_k$ or $h_k$. Then,
\begin{equation}
\frac{\partial \bar{I}_k(i,j)}{\partial v} = \frac{\partial T_k (i,j)}{\partial v} \ast p^{(s)}_{k}(i,j) \ast p^{(d)}(i,j) \ast p^{(m)}_k(i,j),
\end{equation}
\begin{equation}
\text{where }\frac{\partial T_k (i,j)}{\partial v} = \begin{cases}
	1-\tilde{T}_k(i,j), & \text{if $v$ is $l_k$},\\
	\tilde{T}_k(i,j), & \text{if $v$ is $h_k$}.
\end{cases}
\end{equation}

\section*{Acknowledgment}
LLNL-JRNL-760284. 
This work was performed under the auspices of the U.S. Department of Energy by Lawrence Livermore National Laboratory under Contract DE-AC52-07NA27344.
LDRD funding with tracking numbers 16-ERD-006 and 19-ERD-022 were used for this project.
We thank Kyle Champley from LLNL for useful discussions. 
We also thank Markus Baier and Professor Simone Carmignato from University of Padova
for fabricating and providing the star pattern test object utilized in this manuscript.

This document was prepared as an account of work sponsored by an agency of the United States government. Neither the United States government nor Lawrence Livermore National Security, LLC, nor any of their employees makes any warranty, expressed or implied, or assumes any legal liability or responsibility for the accuracy, completeness, or usefulness of any information, apparatus, product, or process disclosed, or represents that its use would not infringe privately owned rights. Reference herein to any specific commercial product, process, or service by trade name, trademark, manufacturer, or otherwise does not necessarily constitute or imply its endorsement, recommendation, or favoring by the United States government or Lawrence Livermore National Security, LLC. The views and opinions of authors expressed herein do not necessarily state or reflect those of the United States government or Lawrence Livermore National Security, LLC, and shall not be used for advertising or product endorsement purposes.


\ifCLASSOPTIONcaptionsoff
  \newpage
\fi



\bibliographystyle{IEEEtran}
%
\vspace{-0.02in}
\bibliography{paper}

%







\end{document}